\documentclass[lettersize,journal]{IEEEtran}
\usepackage{amsmath, amsfonts, amsthm}
\usepackage{algorithm}
\usepackage{algpseudocode} 
\usepackage{array}
\usepackage[caption=false,font=normalsize,labelfont=sf,textfont=sf]{subfig}
\usepackage{textcomp}
\usepackage{stfloats}
\usepackage{url}
\usepackage{verbatim}
\usepackage{graphicx}
\usepackage{cite}
\usepackage[dvipsnames]{xcolor}
\usepackage{graphicx}
\usepackage{subcaption, standalone} 
\usepackage{tikz-qtree}
\usepackage{tikz}
\usepackage{siunitx}
\usepackage{enumitem}
\usepackage{multirow}
\usepackage{booktabs} 

\usepackage{nomencl}
\makenomenclature

\newtheorem{model}{Model} 
\newtheorem{proposition}{Proposition} 

\newcommand{\CSS}{css}
\newcommand{\ES}{es}
\newcommand{\BMS}{bms}

\def\pow{{\rm pow}}

\begin{document}

\title{Mobile Robot Localization Using a Novel Whisker-Like Sensor}
\author{Prasanna K. Routray, Basak Sakcak, Steven M. LaValle, and Manivannan M.
\thanks{P. K. Routray and Manivannan M. are with the Touch Lab, Center for Virtual Reality and Haptics, Indian Institute of Technology Madras, India, 600036 (e-mail: prasanna.routray97@gmail.com, dev.subudhi49@gmail.com, mani@iitm.ac.in)}
\thanks{B. Sakcak and S. M. LaValle are with the Center for Ubiquitous Computing, Faculty of Information Technology and Electrical Engineering, University of Oulu, Finland (e-mail: basak.sakcak@oulu.fi, steven.LaValle@oulu.fi)}} 




\maketitle

\begin{abstract}
Whisker-like touch sensors offer unique advantages for short-range perception in environments where visual and long-range sensing are unreliable, such as confined, cluttered, or low-visibility settings. This paper introduces a framework for contact point estimation and robot localization in a known planar environment using a single whisker sensor. We develop a family of virtual sensor models. Each model maps robot configurations to sensor observations and enables structured reasoning through the concept of \textit{preimages}—the set of robot states consistent with a given observation. The notion of virtual sensor models serves as an abstraction to reason about state uncertainty without dependence on physical implementation. By combining sensor observations with a motion model, we estimate the contact point. Iterative estimation then enables reconstruction of obstacle boundaries. Furthermore, intersecting states inferred from current observations with forward-projected states from previous steps allows accurate robot localization without relying on vision or external systems. The framework supports both deterministic and possibilistic formulations and is validated through simulation and physical experiments using a low-cost, 3D-printed, Hall-effect-based whisker sensor. Results demonstrate accurate contact estimation and localization with errors under \SI{7}{\milli\meter}, demonstrating the potential of whisker-based sensing as a lightweight, adaptable complement to vision-based navigation.
\end{abstract}


\nomenclature[01]{\(L\)}{Total length of the beam or whisker}
\nomenclature[02]{\(a\)}{Load position on undeformed beam, from fixed end, \(0\leq a\leq L\)}
\nomenclature[03]{\(s\)}{arc-length along deformed beam, \(0\leq s\leq L\)}
\nomenclature[04]{\(s_a\)}{arc-length at load position}
\nomenclature[05]{\(\phi(s)\)}{Slope of beam, positive clockwise}
\nomenclature[06]{\(\phi_a\)}{Slope at load point (\(s=s_a\))}
\nomenclature[07]{\(x(s)\)}{Horizontal coordinate of deformed beam}
\nomenclature[08]{\(y(s)\)}{Vertical coordinate of deformed beam}
\nomenclature[09]{\(\delta_{x}\)}{Horizontal displacement at tip, $L-x(s)$}
\nomenclature[10]{\(\delta_{y}\)}{Vertical displacement at tip, $-y(s)$}
\nomenclature[11]{\((p_x, p_y)\)}{Cartesian position of contact point in sensor frame}
\nomenclature[12]{\(\mathcal{P}\)}{Set of contact points in sensor frame}
\nomenclature[13]{\(d\)}{distance between \((p_x, p_y)\) and sensor reference frame}
\nomenclature[14]{\(F\)}{Magnitude of the concentrated point load, applied vertically downward at \(s=s_a\)}
\nomenclature[15]{\(F_x, F_y\)}{Reaction forces at the fixed end of the beam \((s=0)\)}
\nomenclature[16]{\(\kappa\)}{Beam curvature ($\frac{d\phi}{ds}$)}
\nomenclature[17]{\(M(s)\)}{Bending moment at arc-length ($s$)}
\nomenclature[18]{\(E\)}{Young’s modulus of the beam material}
\nomenclature[19]{\(I\)}{Moment of inertia of the beam’s cross-section}
\nomenclature[20]{\(\mathcal{C}\)}{Configuration space}
\nomenclature[21]{\(Q\)}{State space}
\nomenclature[22]{\(Z\)}{Observation space}
\nomenclature[23]{\(h^{-1}(z)\subset Q\)}{Preimage of \(z\in Z\) under \(h:Q\rightarrow Z\)}
\nomenclature[24]{\(h : Q \rightarrow Z\)}{Sensor mapping}
\nomenclature[25]{\(\theta_{add}\)}{Sensor base rotation angle}
\nomenclature[26]{\(T\)}{Homogeneous transform}
\nomenclature[27]{\(t_x\)}{Linear displacement along x-axis}
\nomenclature[28]{\(t_y\)}{Linear displacement along y-axis}
\nomenclature[29]{\(q_r = (q_x, q_y, q_\theta)\)}{State of the robot}
\nomenclature[30]{\(q_s\)}{State of the sensor}
\nomenclature[31]{\(\{x_{E}, y_{E}, \phi_{E}\}\)}{Position and slope of map boundary}
\nomenclature[32]{\(\phi_{E_i}\)}{Instantaneous slope of the boundary from sensor observation}
\nomenclature[33]{\(\gamma\)}{Angle between beam shape and boundary at contact point}
\nomenclature[34]{\(\Delta\)}{Triangulation}
\nomenclature[35]{\(\epsilon\)}{Error in estimation}
\nomenclature[36]{\(\alpha\in [0, 1]\)}{Whisker deflection parameter}
\nomenclature[37]{\(\sigma_{\alpha}\in \mathbb{R}^2\)}{Shape of deflected whisker}
\nomenclature[38]{\(A\subseteq Z\)}{An interval around the noisy observation $z$}
\printnomenclature[1.35cm]

\section{INTRODUCTION}
\IEEEPARstart{W}{hiskers} provide critical sensing functions in animals such as felids, pinnipeds, and rodents, supporting prey detection, fluid-flow perception, and spatial exploration. Rats, for instance, navigate in low-visibility conditions by relying on whiskers to compensate for limited vision~\cite{mitchinson2007feedback}. These biological sensing strategies motivate the use of whisker-inspired sensors in robotic systems operating in environments with limited visibility, clutter, or spatial constraints—conditions where conventional sensors such as cameras and LiDAR often underperform.

Conventional robotic sensing systems primarily rely on vision, LiDAR, and depth sensors, whose performance degrades in close-range interactions, environments with transparent materials, and low-visibility conditions~\cite{chen2022comparative}. In such cases, short-range modalities like bump sensors and whisker sensors offer complementary capabilities. Prior implementations of robotic whiskers have addressed flow detection~\cite{kottapalli2015harbor}, obstacle sensing~\cite{lepora2018tacwhiskers, xu2021triboelectric}, and texture analysis~\cite{evans2009spectral, n2010texture, routray2022towards}, but their use in localization and mapping remains limited. Unlike conventional tactile sensors, whiskers provide both contact and proprioceptive feedback, enabling estimation of contact location, direction, and magnitude from base reaction signals~\cite{huet2017tactile}.

Accurate contact point estimation and robot localization within a known map remain central challenges in whisker-based sensing. Existing approaches typically rely on base moment and reaction forces to infer contact, neglecting additional parameters such as end-slope, normal load, contact coordinates in the sensor frame, and tip deflections. Incorporating this broader set within the virtual sensor framework~\cite{lavalle2012sensing} enables a mathematical abstraction in which sensor mappings associate robot states with observations, and the corresponding preimages characterize uncertainty in the robot state. Iterative filtering under deterministic, possibilistic, or probabilistic formulations can then progressively reduce this uncertainty. Leveraging these richer parameters in virtual sensor models provides more robust strategies for contact estimation, shape reconstruction, mapping, and localization~\cite{bilevich2023sensor}.

This work presents new insights into contact point estimation by introducing virtual sensor models as sensor mappings from the states of an external physical world to a set of observations. This mathematical abstraction enables reasoning directly about the primary source of uncertainty that is given by the preimage of an observation under a particular sensor mapping. A family of virtual sensor models is evaluated in simulation, and the model based on bending moment is further validated through physical experiments. The virtual sensing framework is built on analytical expressions derived from large-angle cantilever beam deflection. Contact point estimates from the bending moment-based model are compared against state-of-the-art methods using physical experiments. In addition, we introduce a low-cost, reconfigurable 3D-printed whisker sensor. Obstacle profiles are estimated using sensor observations and a predefined motion model. For localization, we use temporal observations—derived from a single whisker sensor and known motion—without relying on auxiliary sensing. The approach supports deterministic and possibilistic formulations and can be extended to probabilistic models. It supports both linear and nonlinear motion and sensing dynamics. These capabilities position whisker-based sensing as a viable modality for confined-space robotics, including subterranean, underwater, and cluttered environments where conventional sensing is often impaired.

Section \ref{sec:relatedWork} reviews sensor properties, use cases, existing mathematical models, and filtering techniques. Section \ref{sec:preliminaries} introduces the large-angle cantilever beam model, which forms the basis for the virtual sensor models and their associated preimages. Section \ref{sec:sensorModels} presents the virtual sensor mappings, defines preimages, and explores how they can be composed or combined, including motion modeling. Section \ref{sec:designAndCalibration} describes the sensor design, fabrication, and calibration process. Sections \ref{sec:contactAndShape} and \ref{sec:selfLocalization} present experimental results on contact point estimation, obstacle profile estimation, and localization. Additionally, we assess how orientation sensing contributes to faster reduction of state uncertainty. Finally, we report results that closely match the obstacle profile, contact locations along the whisker, and robot poses.

\section{Background and Related Work}\label{sec:relatedWork}
\subsection{Functional Overview of Whisker Sensors}
Whisker-like sensors offer unique physical and functional properties that make them suitable for robotic perception tasks. These sensors are highly sensitive, capable of detecting forces as small as \SI{4}{\micro\newton}, and able to respond to air or fluid flow disturbances even before physical contact occurs~\cite{deer2019lightweight}. Depending on the material—such as ABS plastic, steel, carbon fiber, or composites—whisker sensors can be fabricated with varying rigidity or flexibility, influencing their mechanical response and sensitivity.

Whiskers operate via mechanical deformation, transduced into electrical signals through changes in resistance~\cite{solomon2006robotic}, capacitance~\cite{stocking2010capacitance}, or magnetic flux density~\cite{kim2019magnetically}. Their typical depth sensing range is short—around \SI{200}{\milli\metre}—making them ideal for close-range obstacle detection and interaction. In robotic applications, they can serve for proprioceptive sensing~\cite{han2016assessing}, providing estimates of the robot’s relative position or motion by analyzing deformation patterns and contact forces.

Due to their compact form factor and passive mechanical design—which allows them to respond mechanically to external forces without requiring active excitation—whisker sensors are suitable for deployment in air or underwater environments. When built with compliant materials such as silicone for waterproofing~\cite{eberhardt2016development}, they can operate reliably in unstructured or harsh environments~\cite{nguyen2022mechanics}. These characteristics enable the use of whisker sensors in tasks such as navigation, obstacle detection, environmental mapping, and localization, particularly in scenarios where conventional range sensors or vision systems are less effective.

\subsection{Applications of Whisker Sensors}
Whisker sensors have been applied across a range of robotic domains due to their high sensitivity to both contact and non-contact stimuli. They are particularly effective for obstacle detection~\cite{kaneko1998active} and navigation~\cite{pearson2007whiskerbot} in low-visibility environments, where visual sensors are limited. Their ability to sense subtle texture variations also makes them valuable for material recognition and surface profiling tasks~\cite{fend2003active, schultz2005multifunctional, solomon2006robotic, routray2022towards}.

In addition to tactile perception, whisker sensors are capable of detecting weak fluid flows, making them suitable for aerial and aquatic robots operating in dynamic environments~\cite{sisanth2017general}. Their responsiveness to soft contact~\cite{lin2022whisker} and vibration enhances real-time feedback during manipulation and grasping tasks, enabling better control of contact forces. Compared to conventional range sensors such as LiDAR, sonar, and depth cameras, whisker sensors provide a complementary modality for short-range perception~\cite{tiwari2022visibility}.

Beyond sensing and mapping, whisker sensors can contribute to haptic feedback systems similar to the work presented in~\cite{okamura2009haptic} and are central to bioinspired robotics, where they replicate natural tactile strategies for environment interaction~\cite{pearson2007whiskerbot, prescott2009whisking, lin2022whisker, wang2023tactile}. Their versatility continues to expand in robotics research focused on perception, exploration, and contact-rich tasks.

\subsection{Mathematical Modeling of Whisker Sensors}\label{sec:mathModels}
\subsubsection{Modeling Whisker Deformation}
\begin{figure}[!ht]
\centering
\subfloat[]{\includegraphics[width=0.23\textwidth]{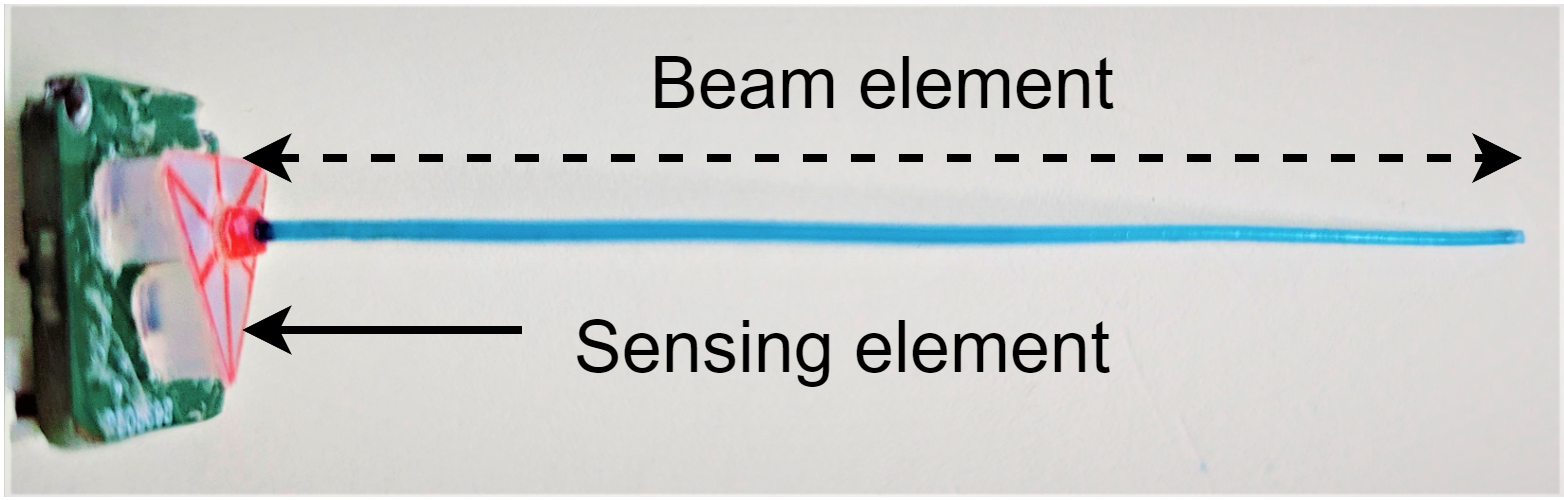}%
\label{fig:sensor-typical}}
\hfil
\subfloat[]{\includegraphics[width=0.25\textwidth, height=3.0cm]{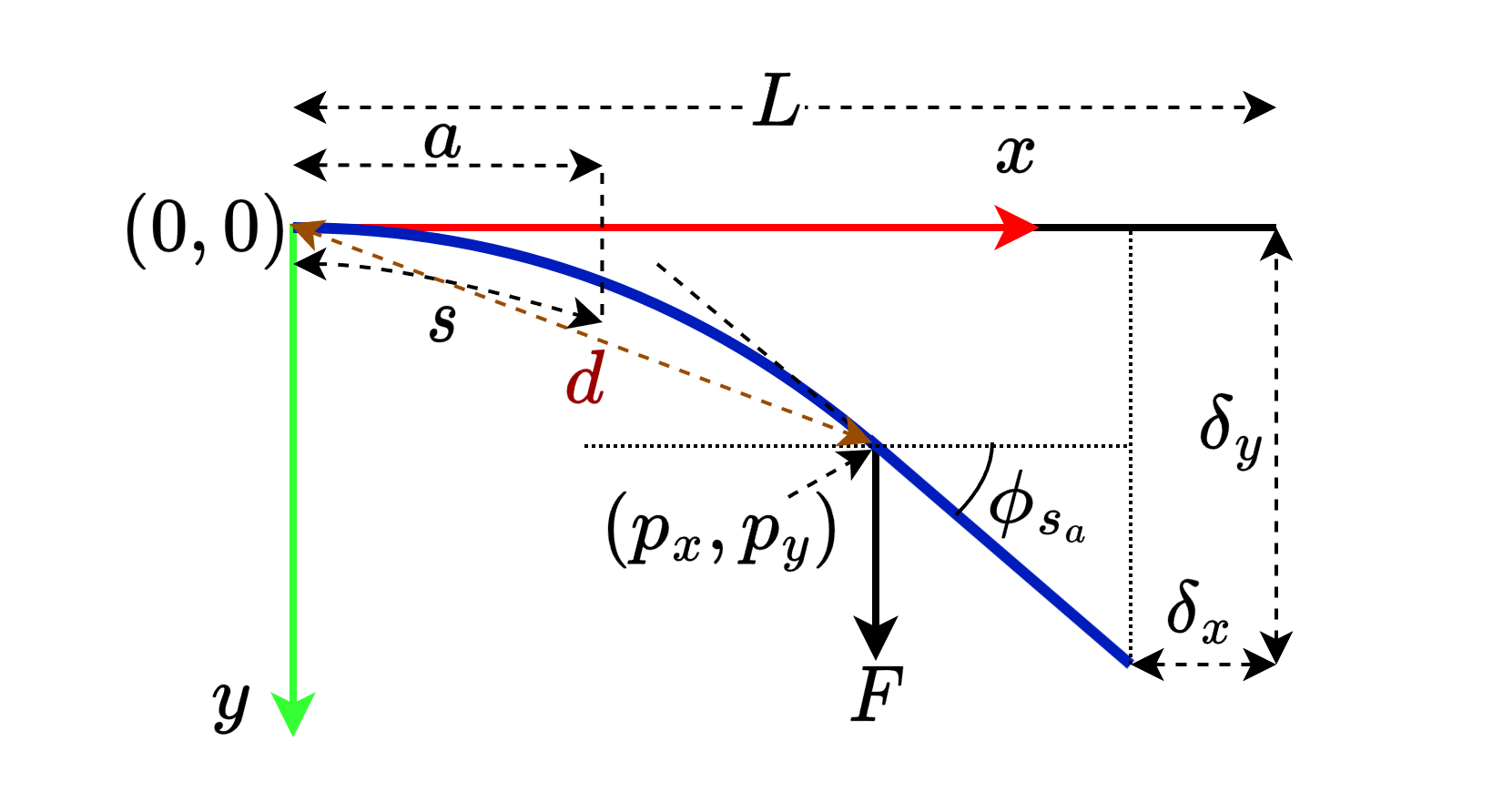}%
\label{fig:beam_LA}}
\caption{Generic representation: (a) A typical physical whisker sensor similar to a cantilever beam. (b) Large-angle deflection of the cantilever beam.}
\label{fig:genericRepresentation}
\end{figure}

Whisker sensors are typically modeled as cantilever beams due to their structural boundary conditions—clamped at the base and free at the tip (see Fig.~\ref{fig:sensor-typical} and Fig.~\ref{fig:beam_LA}). These models approximate whisker deformation under external forces and are foundational for characterizing sensor response. Although biological and synthetic whiskers vary in stiffness, they are generally more rigid than soft materials like elastomers, making them well-suited to beam-based modeling.

Two primary frameworks are used: Euler–Bernoulli beam theory and Cosserat rod theory. Euler–Bernoulli theory assumes small-angle deflections and no shear deformation, and it is widely adopted due to its simplicity and analytical formulation. However, for flexible whiskers or significant tip deflections, large-angle deformation models become necessary. In such cases, the Cosserat rod theory offers a more accurate description, as it incorporates shear, torsion, and complex curvature dynamics~\cite{bauchau2009euler}.

In practice, many whisker sensors operate in regimes where bending is dominant and shear remains minimal. Therefore, large-angle formulations of Euler–Bernoulli theory are often sufficient and computationally efficient. The geometric and material properties of the whisker—including taper, intrinsic curvature, Young’s modulus ($E$), and area moment of inertia ($I$)—significantly affect its static and dynamic behavior. Whiskers with nonlinear taper or curvature often require numerical techniques for accurate modeling~\cite{quist2012mechanical}.

\subsection{Contact Estimation Techniques}
Filtering techniques play a critical role in interpreting whisker sensor observations for tasks such as contact point estimation, obstacle shape estimation, and localization. These methods are generally independent of the specific transduction mechanism, as the physical model—typically based on beam theory—can be decoupled from the transduction process through calibration. This decoupling enables contact point inference from bending moment and measurable quantities such as deflection angle and base whisker motion.

Early work by Kaneko et al.~\cite{kaneko1998active} introduced an active sensing approach in which the whisker was intentionally actuated at the base. Using torque measurements and small-angle beam approximations, they inferred contact distance. Subsequent improvements by Kim et al.~\cite{kim2007biomimetic} measured deflection angles using a pair of sensing elements placed at a fixed distance along the whisker's longitudinal axis. Using Euler–Bernoulli beam theory, this configuration supported contact point estimation followed by object shape discrimination. Solomon et al.~\cite{solomon2008artificial} furthered the research by incorporating lateral slip and Coulomb friction to estimate contact points using base moment. Birdwell et al.~\cite{birdwell2007biomechanical} modeled tapered whiskers and bending moments to estimate radial contact distance.

Advanced techniques have incorporated material properties, intrinsic curvature, and dynamic effects into modeling. Quist et al.~\cite{quist2012mechanical} included intrinsic curvature to estimate base forces and moments. Boubenec et al.~\cite{boubenec2012whisker} demonstrated that vibration frequency shifts, caused by active whisking, correlate with contact location. Iterative approaches, reported by Solomon et al.~\cite{solomon2010extracting}, leverage longitudinal slip for successive contact estimation.

Learning-based techniques have also emerged. Huet et al.~\cite{huet2017tactile, emnett2018novel} used random forests trained on 3D reaction forces and moments at the base of the whisker for contact localization. Lin et al.~\cite{lin2022whisker} applied Bayesian filtering and a soft-contact modeling approach to map environments with compliant objects. More recently, Sofla et al.~\cite{sofla2024haptic} employed temporal filtering over bending moments—measured via a high-resolution ATI Nano sensor—combined with active sensing to estimate the contact point.

Filtering is particularly useful in resolving ambiguities introduced by non-injective sensor mappings. While many methods focus on individual whiskers, some extend to whisker arrays by applying spatial filtering, or to single whiskers by incorporating temporal filtering over time. Despite these developments, few works explicitly characterize the uncertainty in contact estimation or propose unified frameworks that support both deterministic and possibilistic observation models.

\section{Whisker Mechanics and Contact Inference}\label{sec:preliminaries}
\subsection{Brief Review of Beam Theory}\label{sec:review}
We begin with key results from Euler–Bernoulli beam theory, which guide the modeling of whisker sensors and filtering techniques. When a lateral load is applied to a cantilevered beam, it induces a reaction force and a bending moment at the fixed base. The beam undergoes either small-angle deflections (with tip displacements less than 10\% of the total beam length \(L\)) or large-angle deflections, depending on the magnitude of the applied load.

Fig.~\ref{fig:beam_LA} shows the large-angle deflection of a cantilever beam. The deflected tip can be described by displacements \(\delta_x\) and \(\delta_y\), while the beam shape is parameterized by functions \(x(s)\) and \(y(s)\) of arc-length \(s\), with slope \(\phi(s)\) increasing from $0$ to a maximum \(\phi_a\) at the point of load application. For large-angle cases, horizontal tip deflection becomes non-negligible (see \(\delta_x\) in Fig.~\ref{fig:beam_LA}). The curvature along the beam, \(\kappa(s)\), is defined as:
$$\kappa(s) = \frac{d}{ds}\phi(s)  = \frac{M(s)}{EI},$$
and relates the bending moment \(M(s)\) to material properties—Young’s modulus \(E\) and the area moment of inertia \(I\).

For a force \(F\) applied at an arbitrary arc-length \(s_a\), the deflected beam shape (see~\cite{belendez2002large}) for \(0 \leq s \leq s_a\) is given by
\begin{equation}\label{eq:beamShape}
\begin{aligned}
x(s) &= \sqrt{\frac{EI}{2F}} \int_0^{\phi(s)} \frac{\cos \phi}{\sqrt{\sin \phi_a - \sin \phi}} \, d\phi, \\
y(s) &= \sqrt{\frac{EI}{2F}} \int_0^{\phi(s)} \frac{\sin \phi}{\sqrt{\sin \phi_a - \sin \phi}} \, d\phi.
\end{aligned}
\end{equation}
where \(\phi_a = \phi(s_a)\) is the slope at the loading point. If \(\phi_a\) is unknown, it can be numerically estimated from the arc-length using the relation
\begin{equation}\label{eq:beam_Sa}
s_a = \sqrt{\frac{EI}{2F}} \int_0^{\phi_a} \frac{1}{\sqrt{\sin \phi_a - \sin \phi}} \, d\phi.
\end{equation}

For the unloaded section \(s_a < s \leq L\), the beam maintains constant slope \(\phi_a\), and its shape is described by
\begin{equation}\label{eq:beamShape2}
\begin{aligned}
x(s) &= x(s_a) + (s - s_a) \cos \phi_a, \\
y(s) &= y(s_a) + (s - s_a) \sin \phi_a.
\end{aligned}
\end{equation}

The point \((p_x, p_y) = (x(s_a), y(s_a))\) corresponds to the location of contact. Since the integrals involve elliptic functions, closed-form solutions are unavailable, and numerical approximation is required.

In small-angle cases, the horizontal displacement is negligible (\(\delta_x \to 0\)), and the vertical deflection reaches its peak at the free end (\(s = L\)). The bending moment generated at any position \(s < s_a\) from a load \(F\) applied at \(s = s_a\) is given by
\begin{equation}\label{eq:momentCalculation}
\begin{aligned}
M(s) &= F[x(s_a) - x(s)] = F \int_s^{s_a} \cos \phi \, ds,
\end{aligned}
\end{equation}
where \(\cos \phi = \frac{dx}{ds}\). At the base (\(s = 0\)), this simplifies to \(M(0) = F x(s_a)\), and for a load at the tip, \(M(0) = F x(L)\). For \(s > s_a\), \(M(s) = 0\), since the load is applied upstream.

\subsection{Contact Point Estimation via Bending Moment}
Accurately estimating the deflected contact point \((p_x, p_y)\) is key to reconstructing whisker deflection and inferring the geometry of contacted surfaces. Since embedding distributed sensors along a slender elastic beam is often impractical, base measurements (particularly the bending moment \(M(0)\)) are used to infer the location of contact. From a known contact distance along the beam \(s_a\), the full beam shape \((x(s), y(s))\), slope \(\phi(s)\), and tip deflections \((\delta_x, \delta_y)\) can be found using base measurements.

Under small-angle assumptions, Kaneko et al.~\cite{kaneko1998active} proposed estimating contact distance by incrementally actuating the base—either through rotation \(\theta_{add}\)—and measuring the resulting moment \(M(0)\). Solomon et al.~\cite{solomon2010extracting} extended this approach by deriving a closed-form expression for estimating contact distance under linear base motion \(t_y\) as
\begin{equation}\label{eq:distanceEquation}
    d = \sqrt{ \frac{3EI t_y}{M(0)} },
\end{equation}
where \(t_y\) is the displacement of the whisker base along the y-axis, and \(M(0)\) is the corresponding bending moment.

For flexible whisker sensors, small-angle approximations may be inaccurate due to large deflections. In this work, we adopt a large-angle analytical model based on elliptic integrals to define virtual sensor mappings and their preimages. Although analytically insightful, the formulation is computationally expensive and not well suited for real-time use. To address this, we adapt the numerical model developed by Quist et al.~\cite{quist2012mechanical}, which efficiently simulates large-angle 2D deflections. This numerical approach complements the analytical model and serves as the foundation for evaluating virtual sensor behavior and estimating contact points in physical settings.

\section{Virtual Whisker Sensor Models and Preimages}\label{sec:sensorModels}
This section introduces virtual sensor mappings that relate robot states to sensor observations and examines the structure of their preimages—set of states that yield the same observation. We characterize the ambiguity inherent in this mapping and demonstrate how it can be reduced by combining multiple sensor observations.

\begin{figure}[!ht]
\centering
\subfloat[]{\includegraphics[width=0.22\textwidth, height=3.5cm]{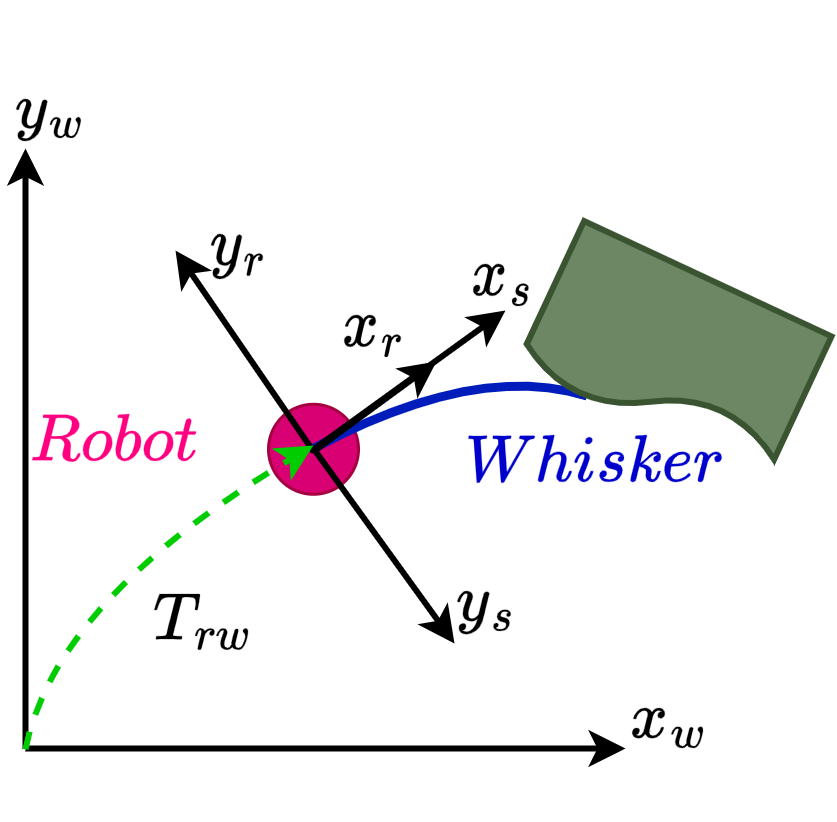}%
\label{fig:robot-arena}}
\hfil
\subfloat[]{\includegraphics[width=0.26\textwidth, height=3.5cm]{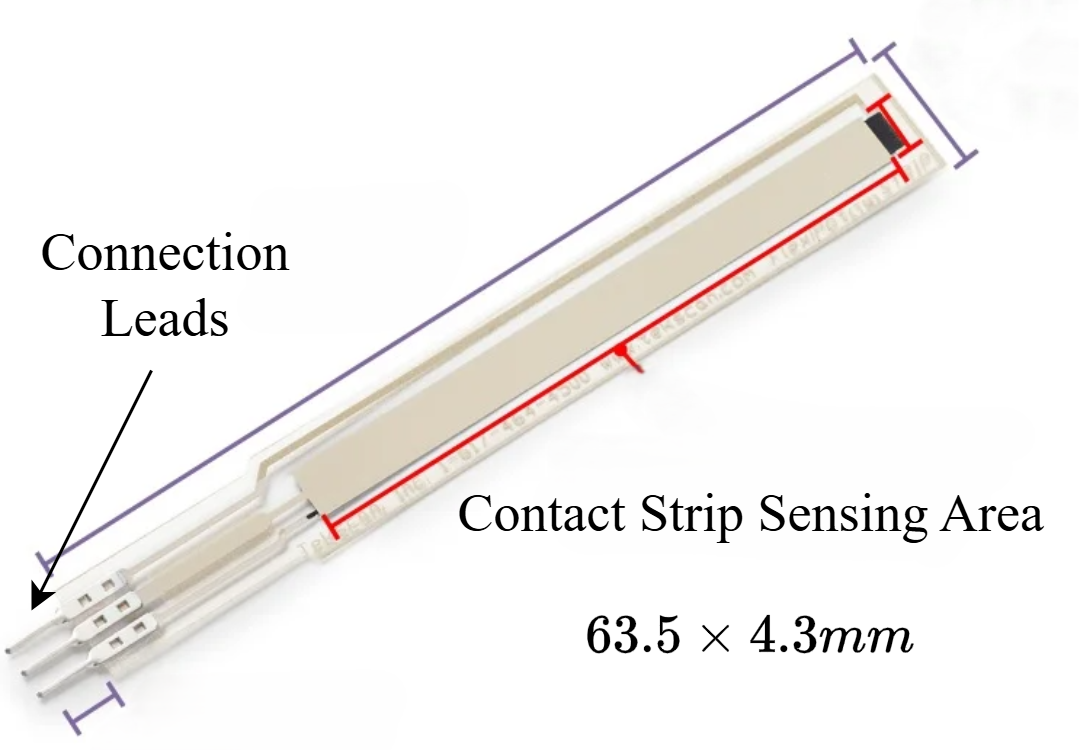}%
\label{fig:contact-strip}}
\caption{Reference frames for the robot \(T_{rw}\) and whisker sensor in world reference frame: (a) Whisker sensor (blue) mounted on a robot and making a contact with the environment. (b) A commercial contact strip sensor available from TekScan\textsuperscript{TM} can act like a whisker sensor.}
\label{fig_OneB}
\end{figure}

\subsection{Virtual Sensor Models}
A \textit{virtual whisker sensor} is a mathematical abstraction designed to simulate tactile sensing without physical hardware. In this study, the virtual sensor is mounted on a mobile robot navigating a planar environment and is represented using geometric transformations: the transform \(T_{sr}\) maps coordinates from the robot frame to the sensor frame, while \(T_{rw}\) maps coordinates from the the world frame to robot frame (see Fig.~\ref{fig:robot-arena}).

The robot state is expressed as the triple \( q = (q_x, q_y, q_\theta)\), in which $(q_x, q_y)$ is the robot position and $q_\theta$ is the orientation with respect to the world frame. The state space $Q \subseteq \mathbb{R}^2\times S^1$, in which $S^1$ is the unit circle, is the set of all the states that the robot can be in.
Previous work has described sensor mapping as a function with domain $Q$ and co-domain $Z$, the set of all possible observations \cite{lavalle2012sensing}. 
However, observations from whisker sensors are inherently ambiguous, since a single robot configuration \(q\) might correspond to multiple feasible whisker shapes and sensor observations \(z \in Z\)~\cite{tiwari2022visibility}. To explicitly handle this ambiguity, we introduce the whisker beam shape parameter \(\alpha \in [\alpha_{\min}, \alpha_{\max}]\), augmenting the original state-space into \(Q_{aug} = Q \times [\alpha_{\min}, \alpha_{\max}]\). Each value of \(\alpha\) defines a specific planar curve \(\sigma_\alpha: [0,1] \to \mathbb{R}^2\) in the sensor frame, parameterized by arc-length \(s\), expressed as coordinates \((x(s), y(s))\). Thus, the mapping from augmented states to sensor observations is given by the function \(h: Q_{aug} \to Z\). This mapping is generally not bijective, as different states may yield the same observation \(z\). We denote by $h^{-1}(z)$, the \emph{preimage} of $z$ under $h$, that is, the set of all robot states and whisker beam shapes that are consistent with the observation $z$.

To effectively analyze this ambiguity in robot states and whisker beam shapes, we introduce a family of \textit{virtual whisker sensor models}, each representing a distinct measurable or non-measurable (load at contact point or end-slope) observation. By combining these sensor models or applying appropriate filters, we can reduce uncertainty in the whisker beam shape and the contact point along the whisker.

\begin{model}[\textbf{Load Sensor}]
\label{model:load-sensor}
This sensor measures the frictionless normal force \( F \) acting laterally at a single point along the whisker beam. Since multiple robot states can lead to distinct beam deformations and consequently different force readings, the mapping is not invertible.

The sensor is defined by
\begin{equation}
    h_{\text{ls}}(q, \alpha) = F,
\end{equation}
where \( q \in Q \) is the robot state, \( \alpha \) is the beam shape, and \( F \in [-F_{\max}, F_{\max}] \) is the sensed lateral force. The maximum magnitude \( F_{\max} \) represents the limit beyond which plastic deformation of beam occurs.

A negative value of \( F \) indicates force applied along the negative \( y_s \)-axis (see Fig.~\ref{fig:beam_LA}). Due to the geometric symmetry of the whisker and sensing mechanism, the sensor response is typically symmetric around zero, though the extreme values may differ in magnitude due to material asymmetries.
\end{model}

\begin{model}[\textbf{Contact Strip Sensor}]\label{model:contact-detectorStrip}
This sensor reports the region of the whisker that is in contact with the environment. Let the whisker be uniformly discretized into intervals \( \mathcal{J} = \{ J_i \}_{i=1}^N \), such that \( \bigcup_{i=1}^{N} J_i = [0, L] \), where each \( J_i = [s_{i-1}, s_i] \) for \( i = 1, \dots, N-1 \), and \( J_N = [s_{N-1}, s_N] \), with \( s_0, s_1, \dots, s_N \) being discretization points along arc-length.

Assuming a single-point contact, the sensor maps the robot state and beam shape to the arc segment that contains the contact point:
\begin{equation}
    h_{\CSS}(q, \alpha) = J_a,
\end{equation}
where \( s_a \in J_a \) denotes the true contact location along the beam.

A typical sensor realization is shown in Fig.~\ref{fig:contact-strip}, where the observation corresponds to the arc interval containing the contact. As shown in Fig.~\ref{fig:beam_LA}, this interval corresponds to \( s_a \) along the whisker.
\end{model}

\begin{model}[\textbf{End-slope sensor}]\label{model:end_slope}
This sensor measures the maximum slope of the whisker beam, denoted by \( \phi_a \), under a transverse normal load. The end-slope sensor is defined by
\begin{equation}
h_{\ES}(q, \alpha) = \phi_a,
\end{equation}
where \( \phi_a \in \left[-\frac{\pi}{2}, \frac{\pi}{2}\right] \) is the angle at the point of contact relative to the undeformed axis (see \( \phi_{s_a} \) in Fig.~\ref{fig:beam_LA}). In practice, the range of valid slopes is limited by physical parameters such as the whisker length \( L \) and material constraints.\\
\end{model}

Given observations from a load sensor, end-slope sensor, or contact strip sensor, the possible set of bending profile of the whisker in the sensor frame can be reconstructed using \eqref{eq:beamShape} and \eqref{eq:beamShape2}. While these models are analytically grounded, most whisker sensor designs rely on base measurements due to constraints on embedding distributed sensors along the slender beam element. To reflect this practical consideration, we introduce an additional virtual sensor model that measures the bending moment at the base—a quantity widely used in engineered whisker sensors and consistent with the biology of whiskers.

\begin{model}[\textbf{Bending moment sensor}]\label{model:moment-sensor}
This sensor quantifies the bending moment at the whisker’s base, modeled as
\begin{equation}
h_{\BMS}(q, \alpha) = M(0),
\end{equation}
where \( M(0) \in [-M_{\text{max}}, M_{\text{max}}] \). A negative (clockwise) moment indicates loading along the \( y_s \)-axis. If the magnitude exceeds \( M_{\text{max}} \), the beam may not return to its original shape.
\end{model}

While individual virtual sensor models provide useful observations—such as load, slope, or contact location—they are generally insufficient to fully determine the robot’s configuration due to the non-invertibility of the sensor mapping. Multiple configurations may lead to identical observations, creating ambiguity in state estimation. In practice, physical whisker sensors often measure only a subset of parameters, such as the bending moment at the base, limiting the available information. To improve accuracy, observations must be combined using motion data (e.g., linear or angular displacement) and filtering methods. The following section characterizes this ambiguity formally using the concept of sensor preimages as discussed in \cite{lavalle2012sensing}.

\subsection{Preimages and Uncertainty}\label{sec:preimage}
The preimage of an observation \( z \in Z \) under a sensor mapping \( h \) defines the set of robot states \( (q, \alpha )\in Q_{aug} \) that could have produced that observation. Given a mapping \( h : Q_{aug} \rightarrow Z \), the state space \( Q_{aug} \) can be partitioned into equivalence classes based on sensor outputs. Two states \( q, q' \in Q \) are indistinguishable under \( h \) if \( h(q, \alpha) = h(q', \alpha) \) for the same beam shape \(\alpha\). Furthermore, the states remain indistinguishable for \( h(q, \alpha) = h(q, \alpha') \) or \( h(q, \alpha) = h(q', \alpha') \). In such cases, no sensor reading can distinguish between them. This section analyzes the preimages for each virtual whisker sensor model, illustrating how ambiguity due to sensing arises and how it varies with sensor type.

\subsubsection{Preimages of a Load Sensor}\label{preimage:load-sensor}
Consider the virtual load sensor defined in Model~\ref{model:load-sensor}, which measures the normal contact force \( F \) applied at an unknown arc-length \( s_a \) along the whisker beam (see Fig.~\ref{fig:beam_LA}). From \eqref{eq:beam_Sa} and \eqref{eq:beamShape}, it is evident that knowing only \( F \) is insufficient to uniquely determine the beam’s shape \( \alpha\), as multiple contact locations can yield the same force.

For different values \( s_{a} \in [0, L] \) and known \( F \), the corresponding end-slopes \( \phi_{a} \) can be computed by solving
\begin{equation}\label{eq:eqnProfileLOAD}
    s_{a} = \sqrt{\frac{EI}{2 F}} \int_0^{\phi} \frac{1}{\sqrt{\sin \phi_{a} - \sin \phi}} \, d\phi,
\end{equation}
where \( \phi \leq \phi_{a} \).

\begin{figure}[!ht]
\centering
\subfloat[]{\includegraphics[width=0.24\textwidth, height=4cm]{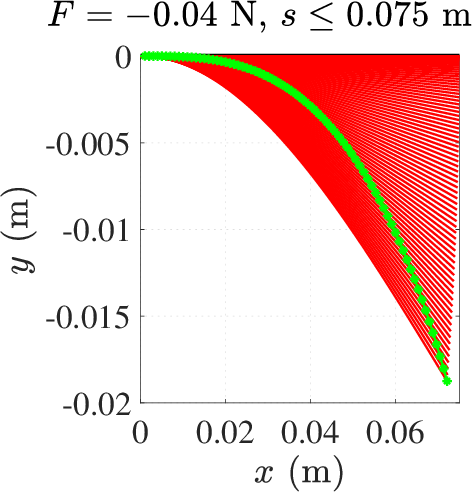}%
\label{fig:loadSensor}}
\hfil
\subfloat[]{\includegraphics[width=0.24\textwidth, height=4cm]{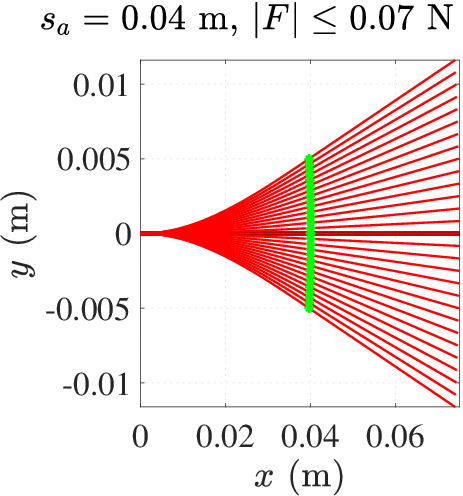}%
\label{fig:ContactStrip}}
\caption{Possible contact locations and corresponding whisker sensor shapes. The green curve represents the set of possible contact locations, while the red curves indicate the potential whisker shapes for each contact point: (a) Load sensor, (b) Contact strip sensor.}
\label{fig_TwoA}
\end{figure}

All end-slopes and corresponding beam shapes satisfying the above equation are illustrated in Fig.~\ref{fig:loadSensor} (red).
Each valid pair \( (s_{a}, \phi_{a}) \) satisfying \eqref{eq:eqnProfileLOAD} generates a unique beam shape \( (x(s), y(s)) \) and corresponding contact point \( (p_x, p_y) = (x(s_a), y(s_a)) \). For computational purposes, \([0, L]\) is discretized, and beam shapes are computed as shown in Fig.~\ref{fig:loadSensor} which also illustrates the associated contact points (green).

This set of contact points, along with their associated beam shapes \( \alpha \), is useful in determining the preimage of a given load sensor observation. Consider a single point along the boundary of the environment where the robot makes contact via the whisker sensor. The robot may occupy different states that yield the same observation \( F \). The distances from these states to the boundary point can be represented as a set where the states are dependent on the beam shapes \( \alpha \).

The preimage can be constructed in two ways: by aligning the different beam shapes with the boundary of the known map (see Section-\ref{sec:constructPreimage}), or by exploiting the direction-dependent distances between the beam origin and the contact point. the direction-dependent distance method essentially reduces to the directional depth sensor presented in \cite{lavalle2012sensing}. Thus, the preimage of a known \( F \) is a subset of the augmented state-space \( Q_{aug} \), defined by
\begin{equation}
    h_{ls}^{-1}(z) = \{ (q, \alpha) \in Q_{aug} \mid h_{ls}(q, \alpha) = F \}.
\end{equation}

\begin{figure}[!ht]
\centering
\subfloat[]{\includegraphics[width=0.24\textwidth, height=4cm]{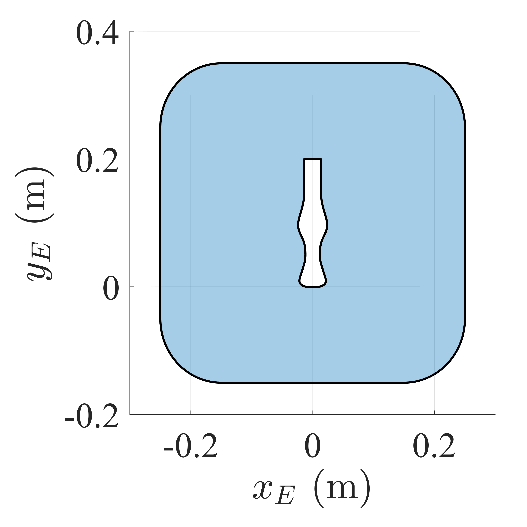}%
\label{fig:mapE}}
\hfil
\subfloat[]{\includegraphics[width=0.24\textwidth, height=4cm]{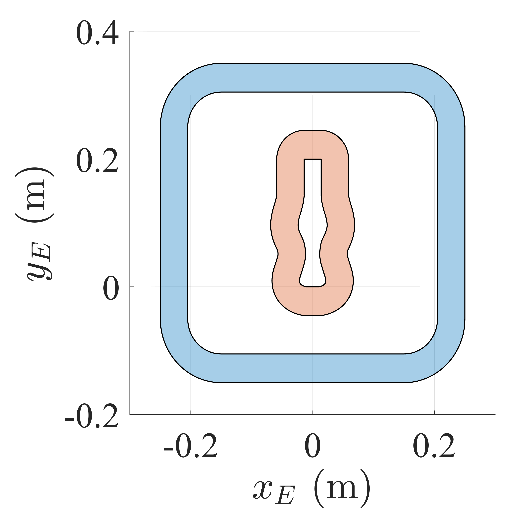}%
\label{fig:depthPreimage}}
\caption{The environment where robot resides and corresponding preimage of $z\in Z$ where $z=F$: (a) A map that features straight, concave, convex, and symmetric boundary sections. The shaded region represents the area where the robot can be located under the assumption that the sensor and robot reference frame are coincident. (b) A single observation reduces the full state-space to the shaded region as the direction-dependent distance determines the width of the strip (shaded region). A particular distance in the set of distances is associated with an orientation of the robot.}
\label{fig:preimageSimple}
\end{figure}

This preimage typically forms a strip-like region around the object boundary (see Fig.~\ref{fig:depthPreimage}), representing the set of robot poses and all possible beam shapes \( \alpha \) consistent with the observed force. The cardinality of the set of \( \alpha \)'s depend on the discretization of \([0, L] \) that is also related to the number of directional distances. This in-turn allows us to have equal number of possible robot states for a single point along the known environment boundary.

\subsubsection{Preimages of a Contact Strip Sensor}\label{preimage:contactStrip-sensor}
Assuming contact at a single point along the whisker, the contact strip sensor measures the arc-length \( s_a \) along the beam. However, when the applied force \( F \) and resulting slope \( \phi_a \) are unknown, the precise contact point and beam shape cannot be uniquely determined, that is we cannot uniquely determine the functions $x(s),y(s)$ and hence, cannot determine the contact point $(p_x,p_y)=(x(s_a),y(s_a))$. Referring to \eqref{eq:beam_Sa}, the set of applied force and end-slope pairs, that is, pairs of $(F, \phi_a)$ that correspond to an observation $s_a$ are the ones that satisfy the equation \eqref{eq:eqnProfileLOAD}. However, the unknown load \( F \in [-F_{\max}, F_{\max}] \), values outside this range cause the beam to undergo plastic deformation.

This leads to a set of beam shapes and contact points as shown in Fig.~\ref{fig:contact-strip}. The preimage \( h_{\CSS}^{-1}(s_a) \) consists of all robot configurations for \((q, \alpha) \in Q_{aug}\) that correspond to a measured contact location \( s_a \). Similar to the previous load sensor case, the preimage can be constructed by aligning the beam shape along the environment boundary or using directional distance that is from the beam origin to the contact point. the possible set of \(\alpha\) and \(q\) for a single point along the environment boundary depends on the discretization of \([-F_{\max}, F_{\max}] \) and geometrically, this possible robot location again resembles to a strip around the environment boundary. In fact, the width of the strip is bounded by the arc-length \( s_a \), as shown in Fig.~\ref{fig:depthPreimage} for all the virtual sensor models. The expression for the preimage set is
\begin{equation}
    h_{\CSS}^{-1}(z) = \{ (q, \alpha) \in Q_{aug} \mid h_{\CSS}(q, \alpha) = s_{a} \}.
\end{equation}

With known $F, \phi_a$, and $s_a$, the beam shape $(x(s), y(s))$ can be uniquely determined by \eqref{eq:beamShape} and \eqref{eq:beamShape2}.

\subsubsection{Preimages of an End-Slope Sensor}\label{preimage:endSlope-sensor}
Given the end-slope \( \phi_a \), we can determine a set of possible contact points and corresponding beam shapes from each valid pair of (\(s_{a}\), \(\sqrt{F}\)) that satisfies the equation \eqref{eq:eqnProfileEndSlopeSensor}. Each contact point results in a distinct deformation profile, and a larger force is required to maintain the same \( \phi_a \) as the contact moves closer to the base. The known end-slope satisfies the relationship
\begin{equation}\label{eq:eqnProfileEndSlopeSensor}
    s_{a} \sqrt{F} = \sqrt{\frac{EI}{2}} \int_0^{\phi_a} \frac{1}{\sqrt{\sin \phi_a - \sin \phi}} \, d\phi
\end{equation}
where \( F \in [-F_{\max}, F_{\max}] \) and \( s_{a} \in [0, L] \) define the range of admissible forces and contact locations.

\begin{figure}[!ht]
\centering
\subfloat[]{\includegraphics[width=0.24\textwidth, height=4cm]{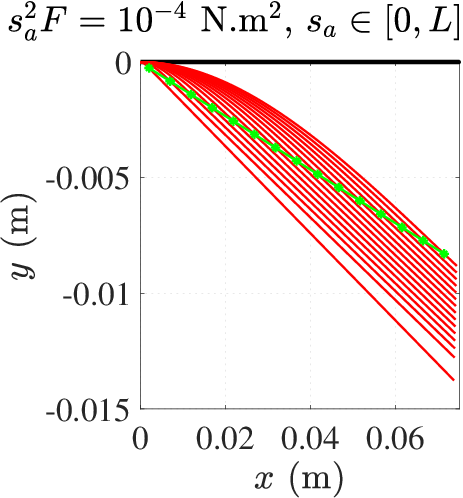}%
\label{fig:endslopeSensor}}
\hfil
\subfloat[]{\includegraphics[width=0.24\textwidth, height=4cm]{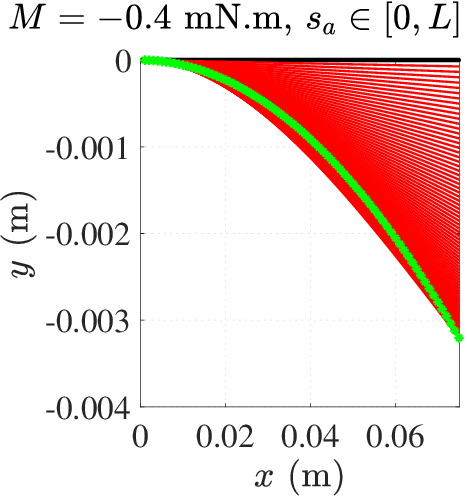}%
\label{fig:momentSensor}}
\caption{Set of possible contact points (green) where the whisker sensor may interact with the obstacle. Corresponding flexible beam shapes for a discrete set of sensor measurements (red): (a) End-slope sensor. (b) Base bending moment sensor in small-angle deflection case.}
\label{fig_TwoB}
\end{figure}

Fig.~\ref{fig:endslopeSensor} shows the set of valid contact points and corresponding beam shapes. The resulting preimage \( h_{\ES}^{-1}(z) \) contains all robot configurations \((q, \alpha) \in Q_{aug}\) consistent with the measured end-slope \( \phi_a \). As with other models (the beam shape alignment or directional distance technique), the preimage is a subset of \(Q_{aug}\) appears as a strip in the known map, illustrated in Fig.~\ref{fig:depthPreimage}. From the sensor observation, both the possible beam shapes and the associated contact points can be derived. Accordingly, the beam shape may be placed using the known end-slope or, alternatively, by applying the directional distance, and in both cases the resulting set of possible robot locations appears as a strip in the known map. The expression for the preimage set is

\begin{equation}
    h_{\ES}^{-1}(z) = \{ (q, \alpha) \in Q_{aug} \mid h_{\ES}(q, \alpha) = \phi_a \}.
\end{equation}

\subsubsection{Preimages of a Bending Moment Sensor}\label{preimage:moment-sensor}

The preimage of a bending moment observation depends on whether the beam undergoes small-angle or large-angle deflection. This can be inferred by comparing the observation to a threshold value. The two cases are analyzed separately.

\paragraph*{Small-Angle Deflection}
In the small-angle regime, where horizontal tip deflection is negligible (\(\delta_x \to 0\)), the bending moment at the base simplifies to a linear relation between force and arc-length. Specifically, the moment is given by
\begin{equation}\label{eq:momentSmallAngle}
M(0) = F s_{a}
\end{equation}
where \(F \in [-F_{\max}, F_{\max}]\) and \(s_{a} \in [0, L]\). This relation allows a set of possible contact points and beam shapes to be computed for a given measured moment using equations \eqref{eq:beam_Sa} and \eqref{eq:beamShape}. The corresponding preimage over the state space \(Q\) is defined as
\begin{equation}
h_{\BMS}^{-1}(z) = \{ (q, \alpha) \in Q_{aug} \mid h_{\BMS}(q, \alpha) = M(0) \}
\end{equation}
and represents all robot states and beam shapes that could result in the same moment.

\paragraph*{Large-Angle Deflection}
In the large-angle case, nonlinear effects become significant, and the moment at the base must be computed using the full beam model. From the expression for base moment in \eqref{eq:momentCalculation}, the moment is given by
\begin{equation}\label{eq:momentLargeAngle}
M(0) = F x(s_{a}) = \sqrt{\frac{F EI}{2}} \int_{0}^{\phi_{a}} \frac{\cos \phi}{\sqrt{\sin \phi - \sin \phi_{a}}} \, d\phi
\end{equation}

This formulation increases the number of unknowns to the tuple \( (F, s_{a}, \phi_{a}) \), where the resulting beam shapes and contact locations must be computed numerically due to the involvement of elliptic integrals. As a result, the \emph{preimage} in the augmented state space \( Q_{\text{aug}} \) is larger than in the small-angle case, because of the nonzero horizontal tip displacement \( \delta_x \), introducing greater uncertainty in the inverse mapping from moment to robot state. However, \emph{geometrically}, the preimage area—considered as a 2D subset of the environment—becomes smaller in the large-angle deflection case, since the increased whisker bending shortens the spatial range of reachable contact points.

\subsection{Contact Uncertainty Reduction via Filtering}\label{sec:filtering}
The preimage obtained from a virtual sensor observation is derived from a set of possible contact points along the whisker. Reducing uncertainty in these contact points and beams shape \( \alpha \) leads to reduced uncertainty in the robot state. To achieve this, we introduce spatial and temporal filtering techniques.

\begin{proposition}
Let \( \mathcal{P} \subset \mathbb{R}^2 \) denote the set of contact points inferred from a virtual whisker sensor model. Suppose \( x(s) \) and \( y(s) \) are continuous functions representing the horizontal and vertical coordinates of the beam shape, parameterized by arc-length \( s \in [0, L] \). Then the set
\begin{equation}\label{eq:contactPointSet}
\mathcal{P} = \left\{ (x(s), y(s)) \mid s \in [0, L] \right\}
\end{equation}
is compact and connected; that is, \( \mathcal{P} \) is bounded and connected.
\end{proposition}

\begin{proof}
The interval \( [0, L] \subset \mathbb{R} \) is closed and bounded, and hence compact and connected. The mapping \( s \to (x(s), y(s)) \) is continuous from \( [0, L] \) into \( \mathbb{R}^2 \). The continuous image of a compact and connected set is compact and connected. Thus, \( \mathcal{P} \subset \mathbb{R}^2 \) is both bounded and connected.
\end{proof}

\begin{proposition}
Let \( \mathcal{P}_1 \) and \( \mathcal{P}_2 \) be two sets of deflected contact points in \( \mathbb{R}^2 \), defined as
\[
\mathcal{P}_i = \left\{ (x_i(s), y_i(s)) \mid s \in [0, L] \right\},
\]
where \( x_i(s) \) and \( y_i(s) \) are continuous functions over the compact interval \( [0, L] \). Then, the intersection set
\[
\mathcal{P}_C := \mathcal{P}_1 \cap \mathcal{P}_2
\]
is compact and connected whenever nonempty. In degenerate cases (singleton or empty), the result holds trivially.
\end{proposition}

\begin{proof}
Each \( \mathcal{P}_i \) is the image of a continuous function from a compact and connected interval \( [0, L] \subset \mathbb{R} \), and therefore \( \mathcal{P}_1, \mathcal{P}_2 \subset \mathbb{R}^2 \) are compact and connected. While the intersection of connected sets need not be connected in general, in this context both sets represent smooth traces of beam deflection generated by physically consistent and continuous models. Any nonempty intersection \( \mathcal{P}_C \) arises from a continuous overlap between these arc-like traces, forming a connected segment or region in \( \mathbb{R}^2 \).

Since the intersection of compact sets is compact, \( \mathcal{P}_C \) is bounded and closed. Thus, \( \mathcal{P}_C \) is connected and bounded whenever nonempty. If \( \mathcal{P}_C \) is a singleton or the empty set, the result is vacuously true.
\end{proof}

The above propositions establish that the sets of contact points obtained from whisker observations are compact and connected, which form the basis for the subsequent spatial and temporal filtering strategies.

\subsubsection{Combining Contact Point Sets from Multiple Sensors}
Sensor fusion improves contact estimation by combining contact point sets inferred from different virtual sensor models. Each model—such as load, end-slope, contact strip, or moment—generates a set of contact points \(\mathcal{P}\) consistent with its observation and the assumed beam model.

\begin{figure}[!ht]
\centering
\subfloat[]{\includegraphics[width=0.24\textwidth, height=4.1cm]{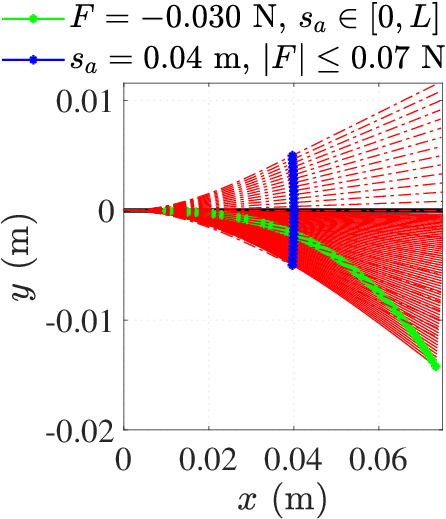}%
\label{fig:loadAndContactSensor}}
\hfil
\subfloat[]{\includegraphics[width=0.24\textwidth, height=3.9cm]{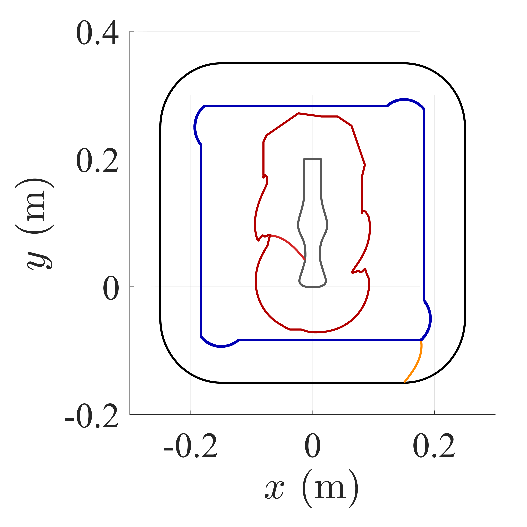}%
\label{fig:combinedSensorPreimage}}
\caption{Fusion of Sensors: (a) Intersection of load and contact strip sensors. Possible whisker configurations (red) in its reference frame for \(s \in [0, L]\)and \(F \in [-F_{max}, F_{max}]\). Possible contact points detected by the load sensor (green) and the contact strip sensor (blue). The intersection of these sets determines the contact point. (b) The reduced uncertainty in contact point results in a preimage which signifies reduced uncertainty in robot state compared to the previous case shown in Fig.~\ref{fig:depthPreimage}.}
\label{fig:combinedSensing}
\end{figure}

These sets can be derived from complementary sensing modalities. Their intersection reduces ambiguity in contact estimation. For example, combining load and contact strip sensor observations gives
\begin{equation}
\mathcal{P}_C = \mathcal{P}_{\text{load}} \cap \mathcal{P}_{\text{strip}}
\end{equation}
where \(\mathcal{P}_C\) denotes the contact points consistent with both sensors. This intersection narrows the set of physically feasible contact locations, as illustrated in Fig.~\ref{fig:combinedSensorPreimage}.

This geometric intersection relates to a set-theoretic formulation over preimages,
\begin{equation}
\Delta(F, s) = h_{ls}^{-1}(F) \cap h_{\CSS}^{-1}(s)
\end{equation}
where \( \Delta(F, s) \) represents the triangulated preimage derived from individual sensor models. This notion of triangulation is adapted from \cite{lavalle2012sensing}. Each observation yields a set of robot states and beam shapes (preimage), and we refer to the intersection of these sets (preimages) as triangulation of observations. This, in-turn reduces uncertainty in robot state.

\subsubsection{Temporal Filtering Using Single Sensor Observations}
Temporal filtering estimates the contact point using a single whisker sensor observations over time. This method assumes that, for small and incremental base motions, the contact location on the object and along the whisker/beam remains unchanged. For larger incremental base motions, the lateral and longitudinal slip plays a significant role and the contact location on the object and along the whisker/beam doesn't remain unchanged. If contact point sets \(\mathcal{P}\) are available at multiple time instants along with base motion information, their intersection provides a filtered estimate of contact point.

Let \( \mathcal{P}_t \) and \( \mathcal{P}_{t+1} \) be the set of contact points at two consecutive time steps. Under the fixed-contact assumption, the intersection of these sets yields a set of new contact points
\begin{equation}\label{eq:getContactPointTemporal}
\mathcal{P}_C = \mathcal{P}_t \cap \mathcal{P}_{t+1}.
\end{equation}

This approach enables contact localization with a single sensor while the robot performs incremental motions. For larger base displacements, however, the whisker may slip across the surface, and the intersection may vanish or shift. This indicates a change in contact and allows slip detection to emerge as a byproduct of the filtering process.

\subsection{Motion Model}\label{sec:motionModel}
The motion model describes how the robot’s configuration evolves over time in response to known control inputs or base displacements. At time \( t \), the robot’s possible configurations consistent with sensor observations form a subset \( \mathbf{Q}_t \subseteq \mathbf{Q} \) obtained from the preimage discrading the beam shapes. Although \( \mathbf{Q}_t \) lies in a continuous space, we discretize it by sampling a finite set \( \hat{\mathbf{Q}}_t = \{(q_{x_k}, q_{y_k}, q_{\theta_k})\}_{k=1}^K \) for computational purposes, where each tuple represents a sampled robot configuration.

To propagate these configurations under known motion, we apply a planar rigid-body transformation defined by a translation \( (t_x, t_y) \) and a rotation \( \theta \). This motion, represented by the transformation matrix \( T_{t} \), is applied relative to each configuration \( q_k \in \hat{\mathbf{Q}}_t \), which is first expressed as a homogeneous transformation matrix \( T_k \).

The updated pose is computed as \( T_k' = T_k \cdot T_{\text{local}} \) when transforms are carried in local frame, and the resulting configuration \( q_k' = (q_{x_k}', q_{y_k}', q_{\theta_k}') \) is extracted by reading the translation and rotation from \( T_k' \). Specifically, the new position is computed by rotating and translating the old position by \( (t_x, t_y) \), and the new orientation is given by \( q_{\theta_k}' = q_{\theta_k} + \theta \). The resulting set \( \hat{\mathbf{Q}}_{t+1} = \{q_k'\}_{k=1}^{K} \) represents the robot’s possible configurations at time \( t+1 \), after applying the known motion model.

\section{Sensor Design and Calibration}\label{sec:designAndCalibration}
Sensor design and fabrication are typically task-specific. The proposed virtual sensors can be partially realized using current technology. Recent advances in materials science and MEMS fabrication have enabled the development of contact strip sensors (see Fig.~\ref{fig:contact-strip}). Among engineered whisker sensors, the bending moment sensor is the simplest and most widely fabricated. Designing a sensor to measure the applied normal load along the whisker beam is highly complex. We developed a simple, low-cost, and reconfigurable whisker sensor that measures magnetic flux density at the base, similar to a bending moment sensor (see Model~\ref{model:moment-sensor}). Its total cost is under \$20~USD, with most parts 3D-printed, making it reconfigurable and accessible for extended use.

\subsection{Sensor Fabrication}\label{sec:sensorFabrication}
\begin{figure}[!htp]
\centering
    \includegraphics[width=\linewidth, height=6cm]{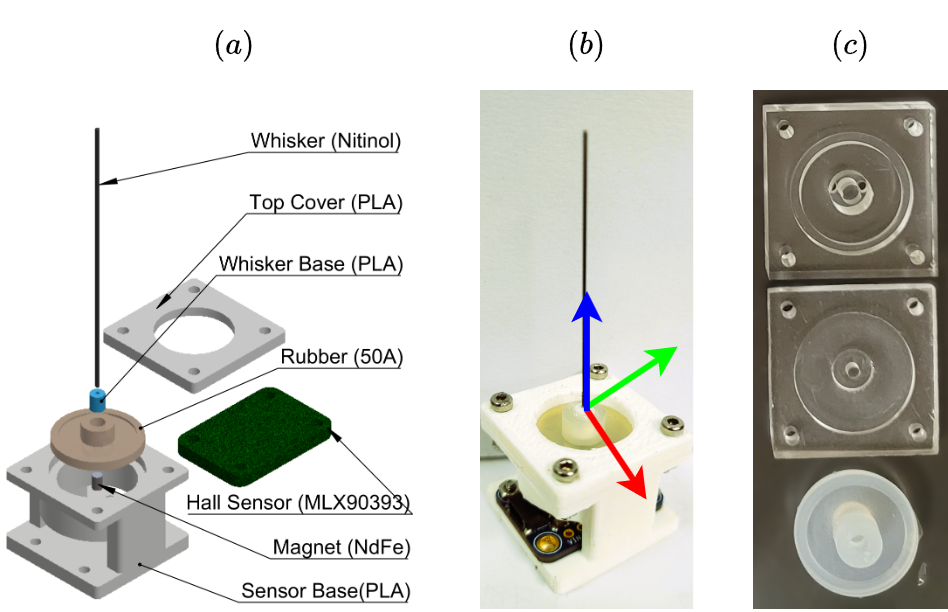}
    \caption{Whisker Sensor: (a) CAD model of the sensor, illustrating all necessary components. (b) Fabricated sensor. (c) Rubber compliant element (CE) and the acrylic mold used for CE fabrication.}
    \label{fig:setupSensor}
\end{figure}
The whisker sensor is designed for modularity and ease of fabrication. Its design is inspired by prior work~\cite{lin2022whisker, kim2019magnetically} and features a straight whisker element, as shown in Fig.~\ref{fig:setupSensor}(a), with the fabricated sensor illustrated in Fig.~\ref{fig:setupSensor}(b). Material selection is guided by prior research to ensure optimal performance. Polylactic acid (PLA) is used to 3D print the sensor base, with a density of \(\rho = \SI{1230}{\kg\per\cubic\meter}\), Young’s modulus \(E = \SI{4.1}{\giga\pascal}\), and Poisson’s ratio \(\nu = 0.35\). A Hall sensor (MLX90393) is used to measure the magnetic response at the base, offering 16-bit resolution across three axes and a range of \SIrange{\pm5}{\pm50}{\milli\tesla}. Alternatively, the TLE493D Hall sensor can be used, providing 12-bit resolution up to \SI{\pm160}{\milli\tesla}. The compliant structure is made from liquid silicone rubber (LSR) with a Shore hardness of 50A, allowing for configurability by varying its thickness and diameter. LSR was prepared by mixing Part-A and Part-B in a 1:1 weight ratio, molded using an acrylic template, and cured for 24 hours. The whisker element is made from a thin Nitinol wire with diameter \(\SI{0.58}{\milli\meter}\), Young’s modulus \(E = \SI{52}{\giga\pascal}\), area moment of inertia \(I = \SI{5.555e-15}{\meter^4}\), and density \(\rho = \SI{6450}{\kg\per\cubic\meter}\), with an effective length of \(\SI{75}{\milli\meter}\).

\subsection{Sensor Calibration}\label{sec:calibration}
The calibration procedure follows~\cite{solomon2010extracting, lin2022whisker}, establishing a mapping between the magnetic flux density \( B \), measured by the Hall sensor, and the bending moment \( M(0) \) at the fixed base of the cantilever beam. Calibration is performed by deflecting the whisker to various contact points \((p_{x_i}, p_{y_i})\) within the YZ-plane and recording corresponding measurements \( (B_i) \), where \( i = 1, 2, \ldots, n \), and \( n = 1000 \) ensures good representation. Similarly, the numerical model\cite{quist2012mechanical} was used to deflect the whisker and record bending moment \( M(0)_i\). The objective is to derive a mapping between \( B_i \) and \( M(0)_i \).

A numerical model simulates the bending of a thin, whisker-like beam under a point load, operating in ``point mode'' based on the formulation by Quist et al.~\cite{quist2012mechanical}. In ``point mode,'' the coordinates of the contact point \((p_x, p_y)\) are known. The model searches for a force \( F \) and arc-length \( s_a \) such that the deflected point matches \((p_x, p_y)\). It then returns the deflected whisker shape and the full 2D components of force and moment at the base.

\begin{figure}
\centering
\subfloat[]{\includegraphics[width=0.24\textwidth, height=4cm]{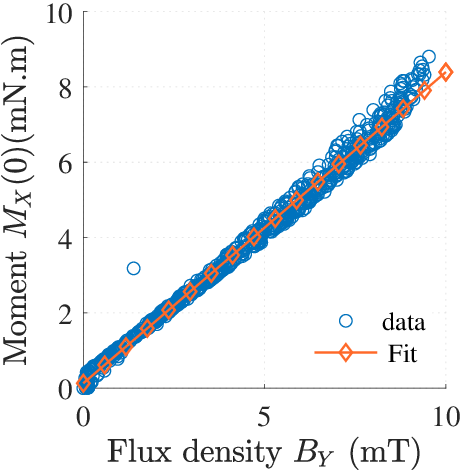}%
\label{fig:FluxToMoment}}
\hfil
\subfloat[]{\includegraphics[width=0.24\textwidth, height=4cm]{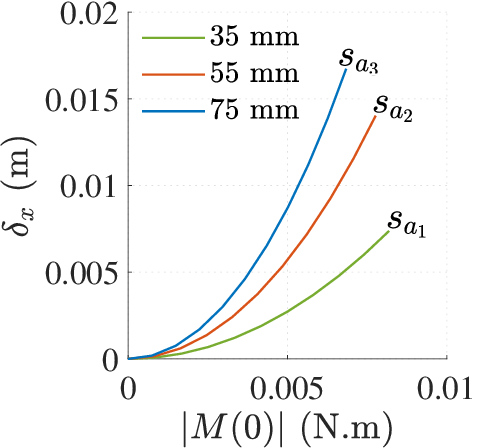}%
\label{fig:MomentToDeltaX}}\centering
\caption{Sensor Calibration: (a) Mapping of flux density to the bending moment at the base. (b) Mapping of bending moment at the base to $\delta_{x}$ (refer to Fig.~\ref{fig:beam_LA} for $\delta_{x}$).}
\label{fig:sensorCalibration}
\end{figure}

All sensor-specific parameters, including geometry and material properties, are consistent with those presented in Section~\ref{sec:sensorFabrication}. To align the simulated response with the real sensor, an elastic boundary condition is applied at the fixed end. The result of this calibration is shown in Fig.~\ref{fig:FluxToMoment}.

For large-angle deflections, the calibration also establishes a relationship between \( M(0) \) and horizontal deflection \( \delta_x \), as described in~\eqref{eq:momentCalculation}. Although this mapping is not one-to-one due to the nonlinear nature of beam deflection, it remains useful. Fig.~\ref{fig:MomentToDeltaX} illustrates this mapping. With rigid whiskers (e.g., carbon fiber), the moment-deflection relation would be one-to-one over a broader range. In contrast, flexible whiskers often exhibit tip slip, especially under higher loads. When contact occurs near the tip, the mappings \( B \rightarrow M(0) \), and \( M(0) \rightarrow \delta_x \), becomes nearly one-to-one and more reliable.

\section{Whisker Contact and Obstacle Shape Analysis}\label{sec:contactAndShape}
The experimental analysis connects the virtual sensor models to two key objectives: (1) contact point determination and (2) obstacle shape estimation. The physical sensor observations do not directly yield the contact point or the robot's state. To extract these, we apply the filtering strategies introduced in Section~\ref{sec:filtering}. Specifically, we use temporal filtering over the set of possible contact points to estimate the location of contact from the outputs of virtual sensor models. The same approach is then applied to the physical whisker sensor to determine the contact point and estimate the obstacle shape. It is important to note that these experiments do not involve explicit motion planning. The contact point and obstacle boundary shape are estimated as the whisker sweeps across the obstacle along a known trajectory.

In experiments, the whisker sensor is mounted on a 3D linear stage with \SI{10}{\micro\metre} resolution. The sensor is connected to a Teensy~4.0 board that streams data at a rate of 300 samples per second. All data acquisition, motion control, and analysis were performed on a Windows 10 system with an Intel Core i5 CPU. MATLAB\textsuperscript{\textregistered} R2023a is used for data analysis, including the \texttt{polyshape}, \texttt{polygon}, and \texttt{polyxpoly} functions for geometric representation and intersection of contact point sets.

\subsection{Experiment: Contact Point Determination}
\begin{figure}[!ht]
\centering
\subfloat[]{\includegraphics[width=0.24\textwidth, height=3.8cm]{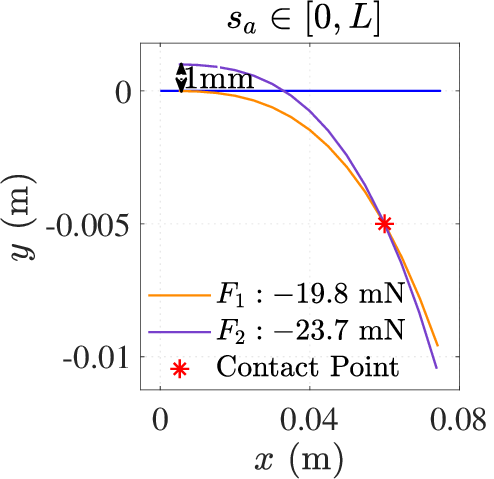}%
\label{fig:contactFromLoadSensor}}
\hfil
\subfloat[]{\includegraphics[width=0.24\textwidth, height=3.8cm]{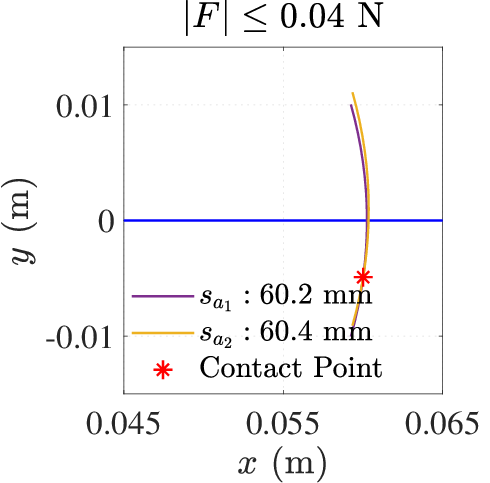}%
\label{fig:contactFromContactStrip}}\centering
\vfil
\subfloat[]{\includegraphics[width=0.24\textwidth, height=3.8cm]{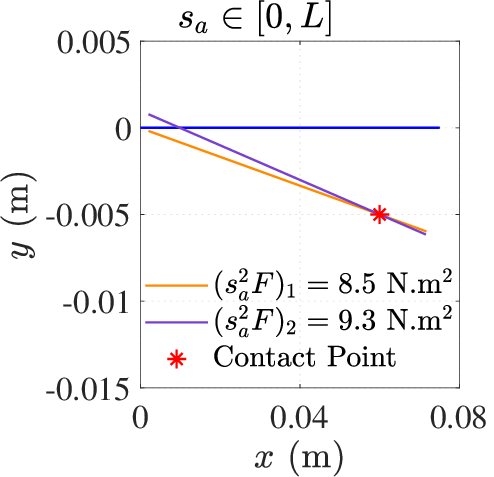}%
\label{fig:contactFromEndslopeSensor}}
\hfil
\subfloat[]{\includegraphics[width=0.24\textwidth, height=3.8cm]{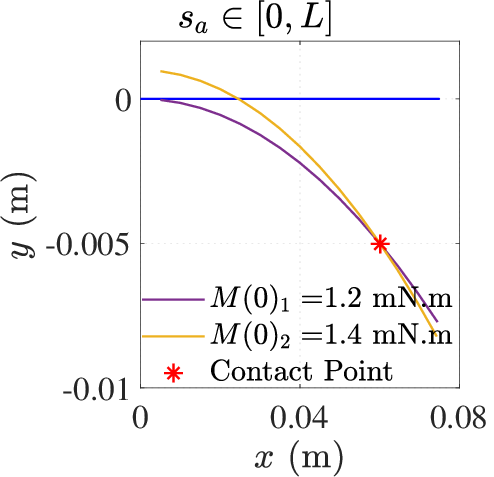}%
\label{fig:contactFromMomentSensor}}
\caption{Contact Point Determination for Simulated Virtual Sensors: The contact point is determined by the intersection of possible contact points using temporal filtering and linear base displacement. (a) Load sensor. (b) Contact strip sensor. (c) End-slope sensor. (d) Moment sensor (small-angle case). A range of parameter values and virtual sensor outputs were considered for contact point estimation.}
\label{fig:contactFromVirtualSensor}
\end{figure}

The contact point along the whisker beam is estimated using the temporal filtering method introduced in Section~\ref{sec:filtering}. This method computes the intersection of the set of possible contact points across consecutive time steps, under the assumption that the contact location remains fixed during small incremental base motions. In our experiments, this approach is implemented using a numerical model adapted from \cite{quist2012mechanical}, which operates in ``force mode" to generate the set of possible contact points. In ``force mode", the experimenter applies a force with known magnitude at a certain arc-length (s) out along the whisker. Under the assumption of no friction, the direction of the force remains normal to the whisker at the point of application. Elastica2D—a numerical simulation tool then computes the shape of the deflected whisker and all 2D components of force and moment at the base. These sets are derived from the deflected beam shapes computed for each sensor observation as shown in Fig.~\ref{fig:loadSensor} and discussed in Section~\ref{preimage:load-sensor}. Fig.~\ref{fig:contactFromVirtualSensor} illustrates the contact points estimated for all virtual sensor models under \SI{1}{\milli\meter} linear base displacement along the $y$-axis.

\begin{figure}
\centering
\subfloat[]{\includegraphics[width=0.24\textwidth, height=3.8cm]{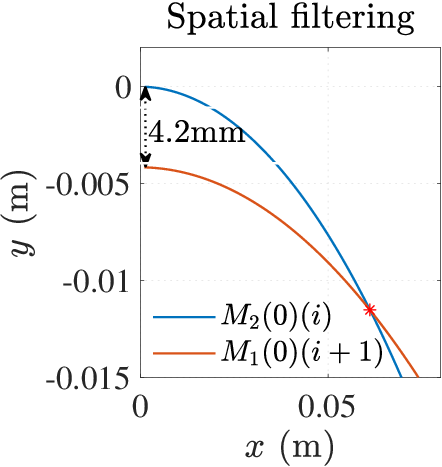}%
\label{fig:SpatialFiltering}}
\hfil
\subfloat[]{\includegraphics[width=0.24\textwidth, height=3.8cm]{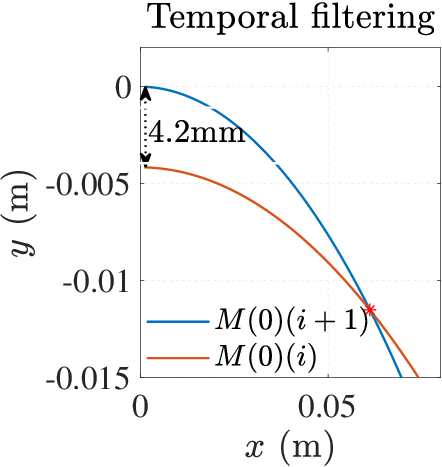}%
\label{fig:TemporalFiltering}}\centering
\caption{Contact Point in Whisker Sensor Reference Frame (Large-Angle Deflection): The point of intersection (Red Star) represents the contact point. Calibration data from Fig.~ \ref{fig:MomentToDeltaX} were used. (a) Spatial filtering can be implemented with two or more sensors placed at predefined positions in the sensor reference frame. $M_1(0,i)$ and $M_2(0, i)$ are recorded by two sensors separated by \SI{4.2}{\milli\meter}. (b) Temporal filtering is applied when a single whisker sensor is in use. The sensor base was moved by \SI{4.2}{\milli\meter}, and $M(0,i) = \SI{17}{\milli\newton\cdot\meter}$, $M(0,i+1) = \SI{25}{\milli\newton\cdot\meter}$
 were recorded.}
\label{fig:filtering}
\end{figure}

The spatial and temporal filtering approach described in Section~\ref{sec:filtering} is applied to physical sensor observations to estimate the contact point. Spatial filtering can also be demonstrated using two physical sensors placed at a fixed distance, both making contact at the same location. Fig.~\ref{fig:filtering} illustrates contact estimation using the physical sensor. The measured magnetic flux density \( B \) is mapped to the bending moment at the base \( M(0) \) using calibration as obtained in Section~\ref{sec:calibration}. The temporal filtering process follows a method similar to that presented in \cite{sofla2024haptic}, and results are compared with the state-of-the-art analytical method described in \cite{kaneko1998active}. Table~\ref{tab:contactResults} presents the contact estimation accuracy. If no intersection between successive sets of possible contact points is found, two scenarios arise: (1) the contact occurs at the tip of the whisker, or (2) contact is lost due to slip. If the sensor output exceeds a predefined threshold and no intersection is found, it confirms contact at the tip.

\begin{table}
\caption{Results of contact point estimation with a physical sensor. The sensor measures incremental moment \( M(0) \) at the base with a base linear displacement of \SI{1}{\milli\metre}. Results are compared with state-of-the-art method~\cite{kaneko1998active, solomon2010extracting}. Here, \( p_x \) and \( \hat{p_x} \) are the true and estimated contact locations, \( d \) is the estimated distance, and \( \epsilon \) is the estimation error.}
\label{tab:contactResults}
\centering
\begin{tabular}{c@{\hskip 5pt}c@{\hskip 5pt}c@{\hskip 7pt}c@{\hskip 5pt}c@{\hskip 7pt}c@{\hskip 5pt}c}
\toprule
\( p_x \) (\si{\milli\metre}) & \( \hat{p_x} \) & \( M_0 \) (\si{\milli\newton\cdot\metre}) & \( \hat{d} \) (Ours) & \( \% \epsilon \) & \( \hat{d} \) (SOTA) & \( \% \epsilon \) \\
\midrule
70 & 71.5 & 0.19 & 71.5 & 2.1 & 67.5 & 3.5 \\
65 & 65.4 & 0.20 & 65.4 & 0.6 & 65.8 & 1.3 \\
55 & 55.8 & 0.28 & 55.8 & 1.4 & 55.6 & 1.1 \\
45 & 44.5 & 0.42 & 44.5 & 1.1 & 45.4 & 0.9 \\
35 & 34.9 & 0.63 & 34.9 & 0.3 & 37.1 & 5.9 \\
\bottomrule
\end{tabular}
\end{table}

\subsection{Experiment: Obstacle Profile Estimation}\label{sec:profileEstimation}
\begin{figure}
\centering
\subfloat[]{\includegraphics[width=0.24\textwidth, height=2.47cm]{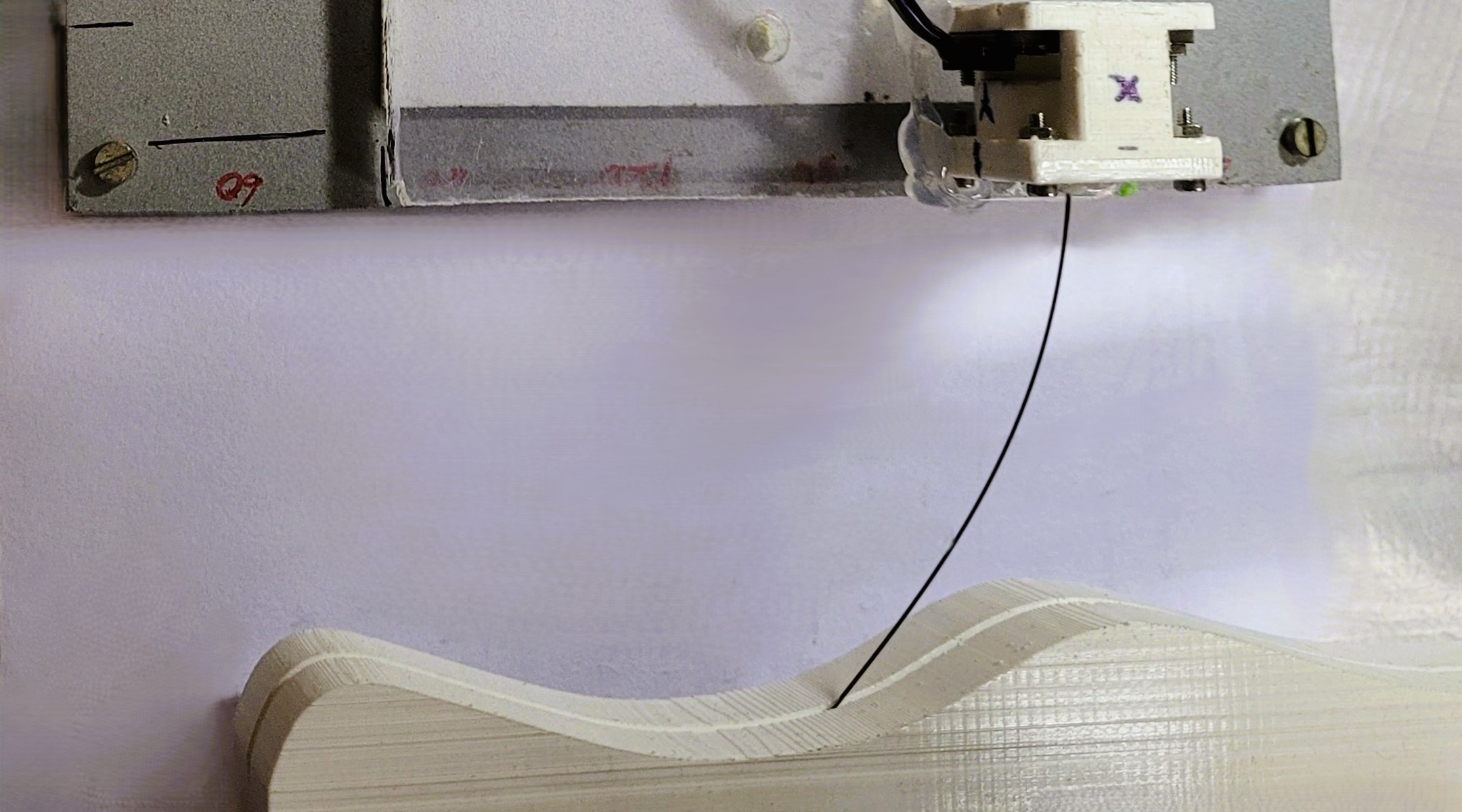}%
\label{fig:expSetupPreImage}}
\subfloat[]{\includegraphics[width=0.20\textwidth, height=2.0cm]{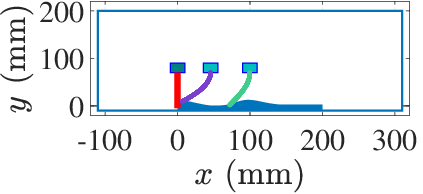}%
\label{fig:sensorAlongBoundary}}
\caption{Experimental and simulation representation of the sensor and obstacle: (a) The whisker sensor (see Fig.~\ref{fig:setupSensor}) is mounted on a linear stage and brushes over an obstacle. (b) The shape of the whisker sensor as estimated.}
\label{fig:expSetup}
\end{figure}

The goal of this experiment is to estimate the shape of an obstacle by sweeping a whisker sensor along its surface and continuously recording contact information. The sensor is mounted on a 3D linear stage with \SI{10}{\micro\meter} resolution; although the stage supports 3D motion, we restrict the sensor to move in the \( yz \)-plane relative to its reference frame (see Fig.~\ref{fig:setupSensor}b). This setup allows controlled base displacements as the sensor interacts with an object. At each time step, the magnetic flux density measured by the physical sensor is converted into a bending moment at the whisker base, which serves as the observation for the virtual moment sensor model. The overall experimental setup is illustrated in Fig.~\ref{fig:expSetupPreImage}, and corresponding simulation results of whisker deflection are shown in Fig.~\ref{fig:sensorAlongBoundary}.

We use the temporal filtering approach introduced in Section~\ref{sec:filtering}, which estimates contact by intersecting the set of possible contact points across successive time steps. This method assumes that contact remains fixed during small base motions, and when the intersection shifts, it reflects slip along the object or whisker. Although slip is a cue for successive contact and shape estimation~\cite{solomon2010extracting}, it is not directly modeled in this work but arises as a by-product. Empirical observations show that, under continued sweeping, contact often shifts toward the whisker tip. This motivates a tip-contact calibration strategy, where the horizontal tip deflection \( \delta_x \) (see Fig.~\ref{fig:beam_LA}) is used to estimate the beam shape when the contact is at the tip. The large-angle deflection formulation is then used to reconstruct the deflected whisker shape and determine the deflected contact point \( (x(s_a), y(s_a)) \) for each sensor observation. These contact points are then accumulated using the known motion trajectory of the whisker base.

\begin{figure}
\centering
\subfloat[]{\includegraphics[width=0.24\textwidth, height=8cm]{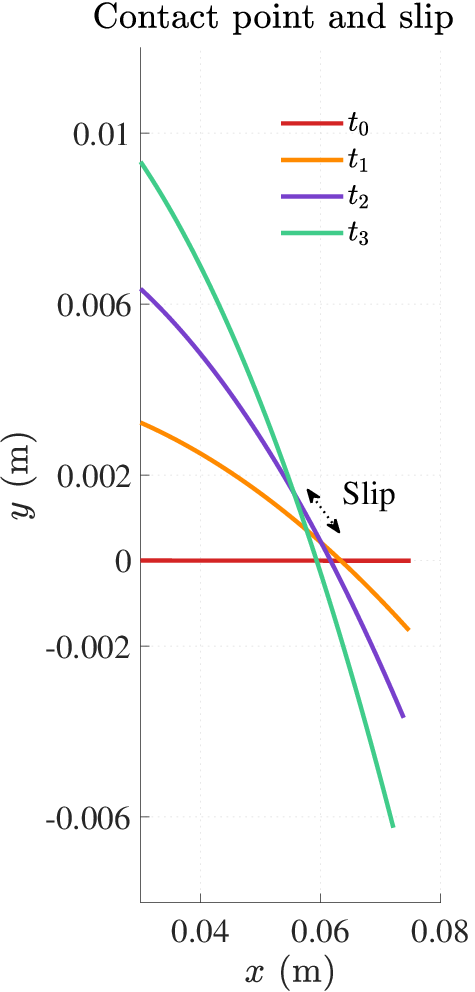}%
\label{fig:continuousContact}}
\hfil
\subfloat[]{\includegraphics[width=0.24\textwidth, height=8cm]{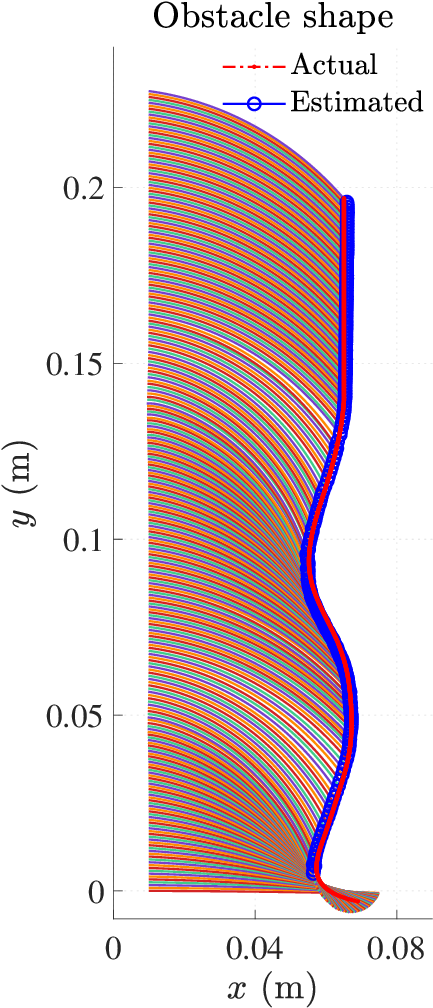}%
\label{fig:profileDetermination}}
\caption{Object Profile Determination: (a) Successive contact points and slip quantification when the base is moved by \SI{4.2}{\milli\meter} in each step. The slip is the shift of contact along the whisker or the boundary while it is sweeping the obstacle boundary. (b) Object profile obtained through successive contact points. The contact point often slips to the tip. While the shape of the whisker sensor at different time instances is shown as estimated, the estimated object profile closely matches the actual profile.}
\label{fig:objectShape}
\end{figure}

To estimate the object boundary, the accumulated contact points across all time steps form a spatial union given by: \(\bigcup_{t=2}^{T_{final}} (x_t(s_a), y_t(s_a))\). This set represents a discrete approximation of the object profile. Slip events can be inferred when consecutive contact points are not coinciding along the whisker, especially under larger base displacements. Fig.~\ref{fig:continuousContact} visualizes these successive contact points and highlights slip transitions, while Fig.~\ref{fig:profileDetermination} shows the estimated object profile constructed from the contact history. For each base motion trajectory, five independent sweeps are performed to generate a reliable profile, which is compared to the ground-truth shape from the 3D CAD model used to fabricate the object.

\section{Robot Localization}\label{sec:selfLocalization}
This section presents a localization framework based on whisker sensor observations, including the inferred contact point and beam shape. We first describe a method to construct the preimage from an observation, followed by the localization procedure. To reduce uncertainty in the robot’s state which is obtained from the preimage, we apply the motion model introduced in Section~\ref{sec:motionModel} for forward projection of known robot states.

\subsection{Constructing a Preimage from Sensor Observation}\label{sec:constructPreimage}
Given a sensor response, the contact point and whisker shape can be used to construct the preimage. A common practical situation arises that the contact point eventually shifts toward the free end of the beam. When the beam shape is known, we align the whisker along the boundary of a known environment map to obtain a set of possible robot configurations. Discretizing the robot's configuration space at any time step \( t \) \( \mathbf{Q}_t \subseteq \mathbf{Q} \) introduces approximation errors, particularly near high-curvature features like sharp corners, where small pose changes can lead to large shifts in contact geometry. This error can cause ambiguity, as several nearby discrete configurations may produce sensor observations that are indistinguishable in the environment. Increasing the sampling density of the configuration space or using a coarse-to-fine search improves localization accuracy while keeping computation efficient.

\begin{figure}[!htpb]
\centering
\subfloat[]{\includegraphics[width=0.45\textwidth]{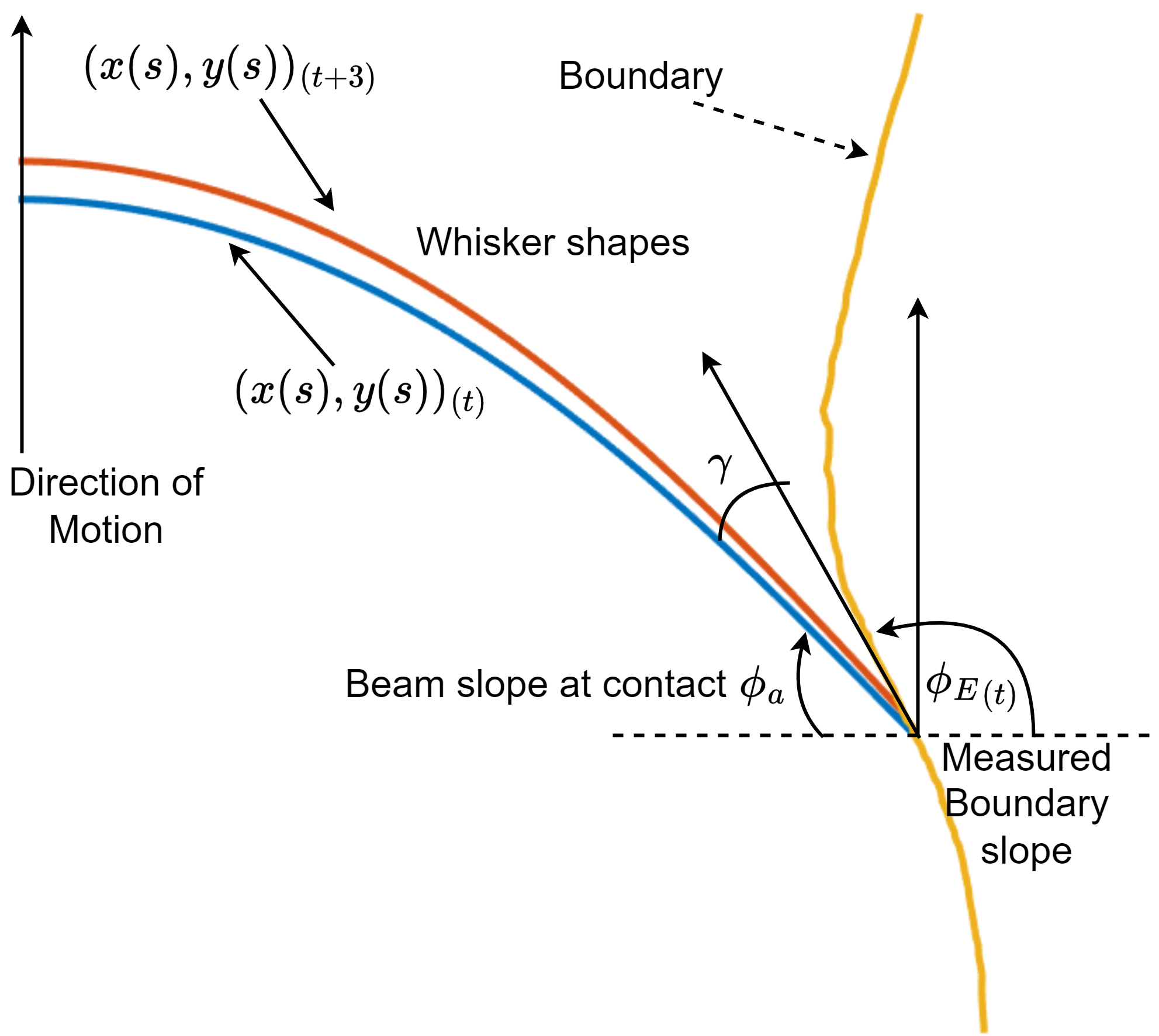}%
\label{fig:TowardsConvex}}
\hfil
\subfloat[]{\includegraphics[width=0.40\textwidth]{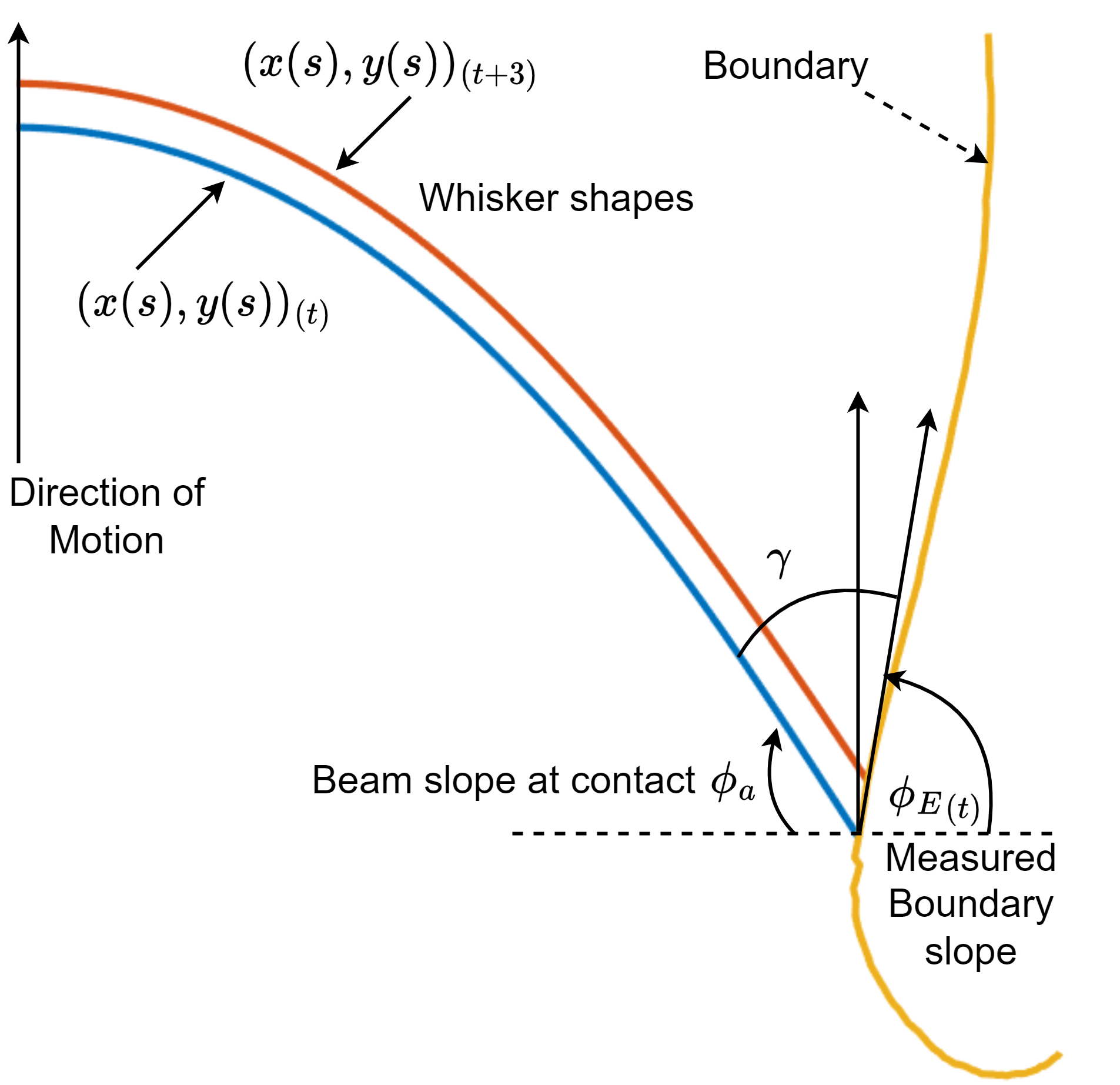}%
\label{fig:awayFromConvex}}\centering
\caption{Whisker contact angle along the environment boundary: a) The sweeping contact approaching a concave region with increasing curvature, b) The whisker contact is moving away from a concave region with decreasing curvature.}
\label{fig:preimageFromObservation}
\end{figure}

As the robot moves, two scenarios may arise: (1) the contact point approaches a concave boundary segment, leading to increasing curvature (see Fig.~\ref{fig:TowardsConvex}); or (2) the contact point moves away from a concave region, resulting in decreasing curvature (see Fig.~\ref{fig:awayFromConvex}). Given the end-slope \( \phi_a \) and the shape of the whisker, we use a short temporal history (2 to 3) of whisker shapes to estimate the instantaneous boundary slope \( \phi_{E_t} \) at the contact point. The instantaneous boundary slope \( \phi_{E_t} \) can be computed using the relative pose derived from the base motion trajectory and whisker shapes observed at time steps \( t \) and \( t+2 \). This estimate is essential for accurately constructing the preimage in the context of a discretized boundary of the environment.

A sensor observation at time index \( t \) corresponds to a specific whisker shape that must be aligned all along the boundary of the environment. The required inclination \( \gamma_t \) between the whisker and the boundary is determined using the whisker slope at contact \( \phi_{a_t} \) and the local boundary slope \( \phi_{E_t} \), obtained from two or more recent whisker samples. The alignment angle \( \gamma_t \) between the whisker and the environment boundary is computed based on the whisker slope at the contact point \( \phi_{a_t} \) and the local boundary slope \( \phi_{E_t} \), with \(\gamma_t = \pi - (\phi_{a_t} + \phi_{E_t})\). Here \( \gamma_t \) represents the angular difference between the whisker and the boundary at the contact point.

With the contact point, whisker shape, boundary slope \( \phi_{E_t} \), and alignment angle \( \gamma_t \) now determined from the sensor observation, all the components required for constructing the preimage are available. These elements form the foundation for computing the robot's pose at any time index \( t \) from preimage by intersecting possible current states and forward projected past states, as described in the following localization method.

\subsection{Localization: Deterministic Method}
To perform self-localization using observations from the whisker sensor, we construct preimages based on the estimated contact point, whisker shape, boundary slope, and alignment angle. The environment is assumed to contain obstacles whose boundaries are defined analytically as continuous functions, which are approximated using discretized points spaced at \SI{1}{\milli\meter} intervals for computational implementation.


At time \( t \), we construct an augmented set \( \mathbf{Q}_{\text{aug}} \) using a minimum of three sensor observations to determine the contact point \( (p_x, p_y) \), whisker orientation \( \gamma_t \), boundary slope \( \phi_{E_t} \), and whisker shape \( x(s), y(s) \). The possible states at time $t$ \( \mathbf{Q}_t \subset \mathbf{Q} \) is obtained by aligning the whisker shape to each discretized point along the known environment boundary such that the inclination angle \( \gamma_t \) is preserved. This alignment determines the robot configuration as
\begin{equation}
q_k = T_k \cdot R(-\phi_a - \gamma_t) \cdot 
\begin{bmatrix}
 -x(s_a) \\
 -y(s_a) \\
 1
\end{bmatrix}
\label{eq:configFromAlignment}
\end{equation}
\noindent
where, \( q_k \in \mathbf{Q}_t \) is a possible robot state obtained by aligning the whisker shape with the environment boundary point represented by the homogeneous transformation \( T_k \in SE(2) \). The contact point \( (x(s_a), y(s_a)) \) is defined in the whisker’s local frame. 
The rotation matrix \( R(\cdot) \) applies the angular correction necessary to align the whisker with the boundary at \( T_k \), and the resulting transformation gives the robot pose consistent with the observed sensor response.

\begin{figure}[!ht]
\centering
\subfloat[]{\includegraphics[width=0.24\textwidth]{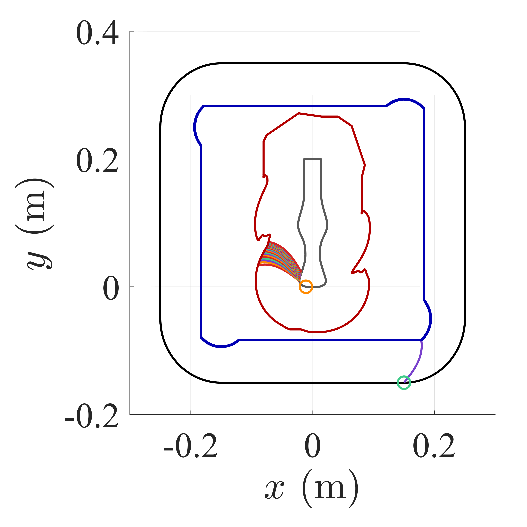}%
\label{fig:manifolds1Reading}}\centering
\hfil
\subfloat[]{\includegraphics[width=0.24\textwidth]{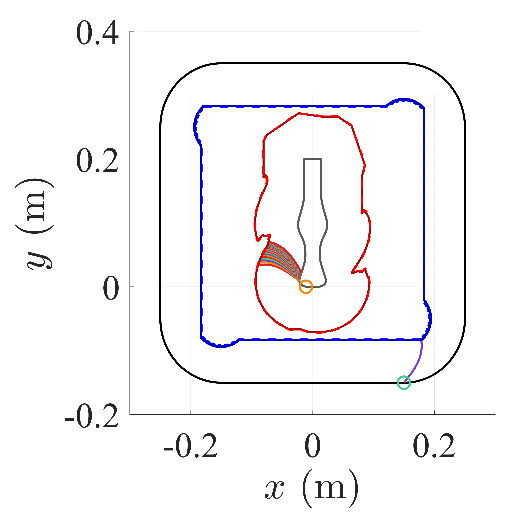}%
\label{fig:manifolds1ReadingMoved}}
\vfil
\subfloat[]{\includegraphics[width=0.24\textwidth]{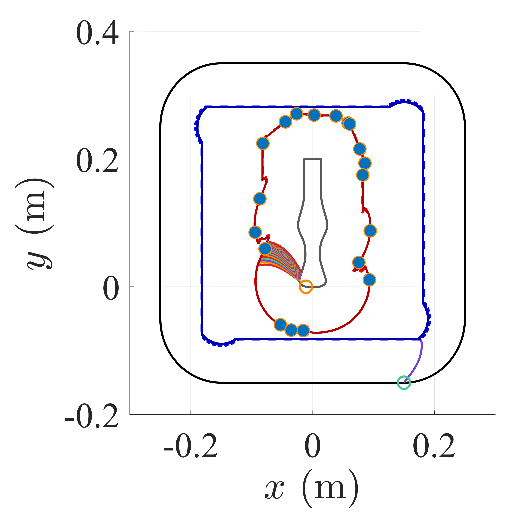}%
\label{fig:manifoldsIntersect1}}
\hfil
\subfloat[]{\includegraphics[width=0.24\textwidth]{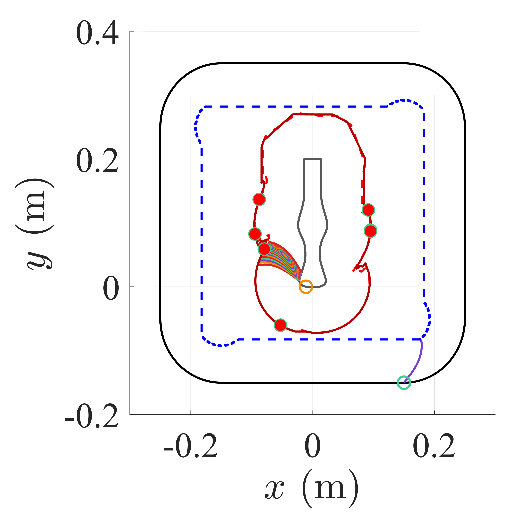}%
\label{fig:manifoldsIntersect2}}\centering
\caption{Deterministic way of localizing the robot: (a) Possible robot states after a sensor observation is recorded (Red and blue). (b) Forward projected robot states (dashed) after motion is applied. (c) Possible states (scattered blue points) of the robot after a new observation is available after motion is applied. (d) Reduced number of robot states after four observations (scattered red points).}
\label{fig:deterministicLocalization}
\end{figure}

Applying the motion model yields the forward-projected set of states \( \hat{\mathbf{Q}}_{t+1} \), obtained by applying the known local transformation \( T_{\text{local}} \) to each state in \( \mathbf{Q}_t \). This discrete set is then converted into a polygonal representation using spline-based interpolation and geometric modeling tools, as shown by the dashed curve in Fig.~\ref{fig:manifolds1ReadingMoved}. A subsequent sensor observation \( z_{t+1} \) produces a new possible set of states \( \mathbf{Q}_{t+1}^{\text{obs}} \) using the same procedure as before. The intersection,
\[
\mathbf{Q}_{t+1} = \hat{\mathbf{Q}}_{t+1} \cap \mathbf{Q}_{t+1}^{\text{obs}}
\] is performed using the \texttt{polyxpoly} operation in MATLAB, yielding a small set of states, as illustrated in Fig.~\ref{fig:manifoldsIntersect1} (19 blue points). This process of sensing, applying motion, and intersection is repeated iteratively, leading to reduced ambiguity in the robot's configuration. A less ambiguous set of states, incorporating additional sensor readings, is shown in Fig.~\ref{fig:manifoldsIntersect2} (6 red points).

Imperfect sensor measurements, calibration errors, and motion model inaccuracies all contribute to uncertainty in the robot’s estimated state. As a result, direct intersection of preimages, as described above, may not yield valid or complete estimates under these imperfections. To address this, possibilistic or probabilistic frameworks can be used to construct and refine the preimage representation. In the following subsection, we describe a possibilistic approach for robot localization that incorporates such uncertainties.

\subsection{Localization: Possibilistic Method}
System-level uncertainties—such as sensor noise, calibration errors, and motion inaccuracies—contribute to the overall imprecision in the robot's estimated state. These sources include errors in sensor calibration (e.g., mapping magnetic flux density to bending moment), sensor noise, the estimation of horizontal tip deflection \( \delta_x \) in nonlinear contact conditions, and imperfections in the motion model. To account for these bounded uncertainties, we adopt a possibilistic framework that incorporates such imperfections into the preimage formulation.

Considering the general case of a set-based noise, the sensor-mapping, denoted by $H$, is defined to take noise into account so that 
\begin{equation}
    H: Q\times \mathcal{A} \rightarrow \pow(Z),
\end{equation}
in which $\mathcal{A}=[\alpha_{\text{min}}, \alpha_{\text{max}}]$, $Z$ is the observation space, and $\pow(\cdot)$ denotes the power set. Let \( \varepsilon > 0 \) be the known bound on sensor observation noise. Specifically, the sensor-mapping we use is defined according to the following equation 
\begin{equation}
    H(q, \alpha) = \{ z\in Z \mid |h(q, \alpha) - z| \leq \varepsilon \},
\end{equation}
in which $h(\cdot)$ is the sensor-mapping considering the deterministic setting. Consequently, given an observation $z$ the set of possible states that could have produced $z$ under $H$ is
\begin{equation}
    H^{-1}(z) = \{ ( q \in Q,\; \alpha \in \mathcal{A}) \mid z \in H(q,\alpha) \}.
\end{equation}
This is akin to the notion of a function's preimage. In our implementation of localization under a possibilistic framework, uncertainty is set to \( \varepsilon = 5\% \) of the observed value, based on sensor characteristics.

\begin{figure}[!ht]
\centering
\subfloat[]{\includegraphics[width=0.24\textwidth]{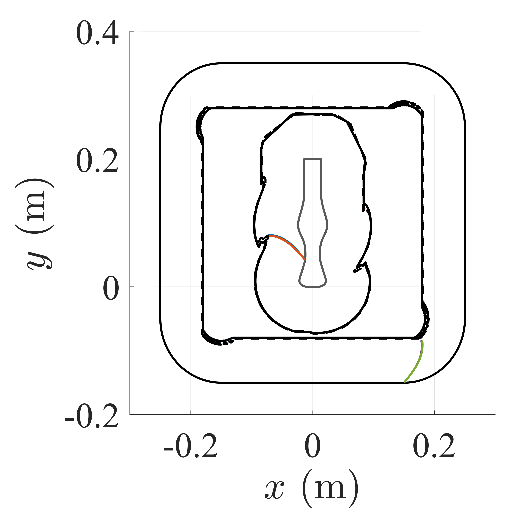}%
\label{fig:patches2Reading}}
\hfil
\subfloat[]{\includegraphics[width=0.24\textwidth]{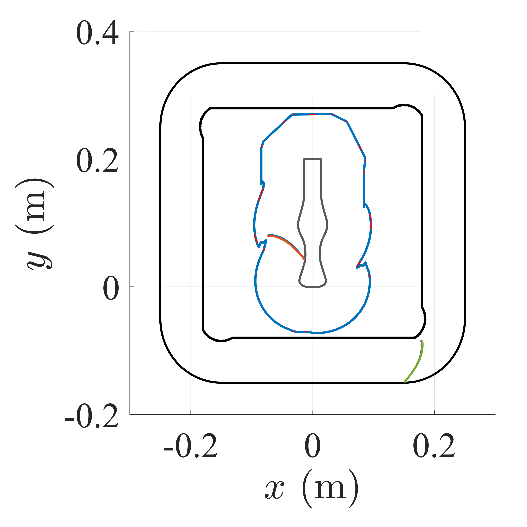}%
\label{fig:patches2ReadingIntersect}}\centering
\caption{Possibilistic Way of Localizing the Robot with Sensing and Motion Model Imperfections: (a) Preimage obtained with the second measurement. (b) Possible subsets (Red) of $Q_{PM_{i+1}}$ where the robot could be located after the third measurement.}
\label{fig:possibilisticLocalization}
\end{figure}

Figure~\ref{fig:possibilisticLocalization} illustrates the possibilistic approach to robot localization under sensing and motion model uncertainties. In Fig.~\ref{fig:patches2Reading}, the preimage is shown after incorporating the second measurement, taking into account bounded uncertainty in the sensor response. This preimage represents a region of the state space where the robot could plausibly be located, rather than a precise point. After the robot undergoes a motion update and a third observation is obtained, the forward projected set of states \( \mathbf{Q}_{t+1} \) is intersected with the possibilistic configuration obtained from preimage of a new sensor observation. As shown in Fig.~\ref{fig:patches2ReadingIntersect}, this intersection yields refined subsets (highlighted in red) representing more likely robot configurations. This iterative process allows the localization method to progressively narrow down the robot stated in the presence of component-level uncertainties.

\subsection{Localization Error Analysis}
To evaluate localization accuracy, we recorded the ground-truth position of sensor from the linear stage (with \SI{10}{\micro\meter} resolution), along with corresponding sensor observations. The preimage-based localization method performs reliably as the sensor sweeps along concave boundaries and produces consistent variations in observation. In deterministic localization, the average localization error remains below \SI{7}{\milli\meter}. Experimental sweeps were conducted for different sweep IDs under similar conditions, as shown in Fig.~\ref{fig:expSetup}. Each sweep ID corresponds to a known location from which obstacle sweeping begins, and consequently, where the localization process starts. We varied the motion model parameters \((t_x, t_y)\), including the number of samples used to estimate the instantaneous boundary slope \( \phi_{E_i} \), typically ranging from 2 to 5. The estimation of \( \phi_{E_i} \) plays a critical role in constructing accurate preimages, particularly when the map boundary is a discrete approximation. To maintain high accuracy, the linear motion per step (\( t_x \) or \( t_y \)) was constrained to less than \SI{1}{\milli\metre}, consistent with the spatial resolution of the discretized environment boundary.

\begin{table}[htb]
\caption{Average localization error for different whisker sweep ID along the obstacle boundary when the contact is known to be at the tip.}
\label{tab:localizationError}
\centering
\begin{tabular}{lccccc}
\toprule
\textbf{Sweep ID} & 01 & 02 & 03 & 04 & 05 \\
\midrule
Average Error \( \epsilon \) (\si{\milli\metre}) & 2.79 & 6.20 & 3.60 & 5.50 & 7.30 \\
\midrule
Standard Deviation \( \sigma \) (\si{\milli\metre}) & 1.09 & 1.67 & 0.02 & 0.65 & 0.44 \\
\bottomrule
\end{tabular}
\end{table}

\section{Conclusion and Future Work}
This paper presented a method for contact point estimation and preimage-based robot localization using a whisker sensor. By leveraging the deflected shape of the whisker beam and the estimated contact point, the robot’s pose was estimated through the intersection of numerically obtained sets of possible states. Unlike traditional approaches that rely on analytical or probabilistic models, the proposed framework uses discrete representations of the environment boundary and beam shape to localize the sensor or robot within a known environment. The evaluation through mathematical models and visualizations demonstrated consistent localization performance, particularly in concave obstacle boundaries.  

Accurate localization, however, requires careful calibration, since mechanical misalignment or modeling inaccuracies can degrade the quality of sensor observations. The inherent nonlinearity of the whisker sensor complicates precise contact estimation, although the numerical model by Quist et al.~\cite{quist2012mechanical} enables mapping between physical readings and the resulting forces and moments at the whisker base. These challenges highlight the need for structured computational strategies, particularly when extending the framework to higher-dimensional state spaces.

Future work will extend the approach in several directions. Building on the deterministic and possibilistic models, probabilistic methods will be investigated to fuse information from multiple sensing modalities. Robustness can be improved by combining whisker observations with long-range sensing, while integrating MEMS-based IMUs will provide orientation information to reduce ambiguity and accelerate localization. Extending the approach to three-dimensional Cartesian spaces and cluttered planar environments will require optimal motion strategies, refined intersections of possible states, and efficient real-time management of iterative forward projections.

\section*{Acknowledgments}
The authors gratefully acknowledge Prof. Mitra J. Hartmann for providing access to the \textit{\textbf{elastica2D}} model, which was instrumental in this study.
\clearpage
\bibliographystyle{./IEEEtran}
\bibliography{./bibliography} 

@article{belendez2002large,
  title={Large and small deflections of a cantilever beam},
  author={Bel{\'e}ndez, Tarsicio and Neipp, Cristian and Bel{\'e}ndez, Augusto},
  journal={European journal of physics},
  volume={23},
  number={3},
  pages={371},
  year={2002},
  publisher={IOP Publishing}
}

@book{lavalle2012sensing,
  title={Sensing and filtering: A fresh perspective based on preimages and information spaces},
  author={LaValle, Steven Michael},
  year={2012},
  publisher={Now Publishers}
}

@inproceedings{tiwari2022visibility,
  title={Visibility-inspired models of touch sensors for navigation},
  author={Tiwari, Kshitij and Sakcak, Basak and Routray, Prasanna and Manivannan, M and LaValle, Steven M},
  booktitle={2022 IEEE/RSJ International Conference on Intelligent Robots and Systems (IROS)},
  pages={13151--13158},
  year={2022},
  organization={IEEE}
}

@inproceedings{routray2022towards,
  title={Towards multidimensional textural perception and classification through whisker},
  author={Routray, Prasanna Kumar and Kanade, Aditya Sanjiv and Tiwari, Kshitij and Pounds, Pauline and Muniyandi, Manivannan},
  booktitle={2022 IEEE International Symposium on Robotic and Sensors Environments (ROSE)},
  pages={1--7},
  year={2022},
  organization={IEEE}
}

@article{kaneko1998active,
  title={Active antenna for contact sensing},
  author={Kaneko, Makoto and Kanayama, Naoki and Tsuji, Toshio},
  journal={IEEE Transactions on robotics and automation},
  volume={14},
  number={2},
  pages={278--291},
  year={1998},
  publisher={IEEE}
}

@article{birdwell2007biomechanical,
  title={Biomechanical models for radial distance determination by the rat vibrissal system},
  author={Birdwell, J Alexander and Solomon, Joseph H and Thajchayapong, Montakan and Taylor, Michael A and Cheely, Matthew and Towal, R Blythe and Conradt, Jorg and Hartmann, Mitra JZ},
  journal={Journal of neurophysiology},
  volume={98},
  number={4},
  pages={2439--2455},
  year={2007},
  publisher={American Physiological Society}
}

@article{deer2019lightweight,
  title={Lightweight whiskers for contact, pre-contact, and fluid velocity sensing},
  author={Deer, William and Pounds, Pauline EI},
  journal={IEEE Robotics and Automation Letters},
  volume={4},
  number={2},
  pages={1978--1984},
  year={2019},
  publisher={IEEE}
}

@inproceedings{lin2022whisker,
  title={Whisker-Inspired Tactile Sensing for Contact Localization on Robot Manipulators},
  author={Lin, Michael A and Reyes, Emilio and Bohg, Jeannette and Cutkosky, Mark R},
  booktitle={2022 IEEE/RSJ International Conference on Intelligent Robots and Systems (IROS)},
  pages={7817--7824},
  year={2022},
  organization={IEEE}
}

@article{solomon2010extracting,
author = {Joseph H. Solomon and Mitra J. Z. Hartmann},
title ={Extracting Object Contours with the Sweep of a Robotic Whisker Using Torque Information},
journal = {The International Journal of Robotics Research},
volume = {29},
number = {9},
pages = {1233-1245},
year = {2010},
doi = {10.1177/0278364908104468}
}

@inproceedings{bilevich2023sensor,
  title={Sensor Localization by Few Distance Measurements via the Intersection of Implicit Manifolds},
  author={Bilevich, Michael M and LaValle, Steven M and Halperin, Dan},
  booktitle={2023 IEEE International Conference on Robotics and Automation (ICRA)},
  pages={1912--1918},
  year={2023},
  organization={IEEE}
}

@incollection{sisanth2017general,
  title={General introduction to rubber compounding},
  author={Sisanth, KS and Thomas, MG and Abraham, J and Thomas, S},
  booktitle={Progress in rubber nanocomposites},
  pages={1--39},
  year={2017},
  publisher={Elsevier}
}

@inproceedings{kim2019magnetically,
  title={A magnetically transduced whisker for angular displacement and moment sensing},
  author={Kim, Suhan and Velez, Camilo and Patel, Dinesh K and Bergbreiter, Sarah},
  booktitle={2019 IEEE/RSJ International Conference on Intelligent Robots and Systems (IROS)},
  pages={665--671},
  year={2019},
  organization={IEEE}
}

@article{kim2007biomimetic,
  title={Biomimetic whiskers for shape recognition},
  author={Kim, DaeEun and M{\"o}ller, Ralf},
  journal={Robotics and Autonomous Systems},
  volume={55},
  number={3},
  pages={229--243},
  year={2007},
  publisher={Elsevier}
}

@article{solomon2006robotic,
  title={Robotic whiskers used to sense features},
  author={Solomon, Joseph H and Hartmann, Mitra J},
  journal={Nature},
  volume={443},
  number={7111},
  pages={525--525},
  year={2006},
  publisher={Nature Publishing Group UK London}
}

@inproceedings{kottapalli2015harbor,
  title={Harbor seal whisker inspired flow sensors to reduce vortex-induced vibrations},
  author={Kottapalli, AGP and Asadnia, Mohsen and Miao, JM and Triantafyllou, MS},
  booktitle={2015 28th IEEE International Conference on Micro Electro Mechanical Systems (MEMS)},
  pages={889--892},
  year={2015},
  organization={IEEE}
}

@inproceedings{schultz2005multifunctional,
  title={Multifunctional whisker arrays for distance detection, terrain mapping, and object feature extraction},
  author={Schultz, Aimee E and Solomon, Joseph H and Peshkin, Michael A and Hartmann, Mitra J},
  booktitle={Proceedings of the 2005 IEEE International Conference on Robotics and Automation},
  pages={2588--2593},
  year={2005},
  organization={IEEE}
}

@article{pearson2007whiskerbot,
  title={Whiskerbot: a robotic active touch system modeled on the rat whisker sensory system},
  author={Pearson, Martin J and Pipe, Anthony G and Melhuish, Chris and Mitchinson, Ben and Prescott, Tony J},
  journal={Adaptive Behavior},
  volume={15},
  number={3},
  pages={223--240},
  year={2007},
  publisher={Sage Publications Sage UK: London, England}
}

@article{solomon2008artificial,
  title={Artificial whiskers suitable for array implementation: accounting for lateral slip and surface friction},
  author={Solomon, Joseph H and Hartmann, Mitra JZ},
  journal={IEEE Transactions on Robotics},
  volume={24},
  number={5},
  pages={1157--1167},
  year={2008},
  publisher={IEEE}
}

@inproceedings{fend2003active,
  title={An active artificial whisker array for texture discrimination},
  author={Fend, Miriam and Bovet, Simon and Yokoi, Hiroshi and Pfeifer, Rolf},
  booktitle={Proceedings 2003 IEEE/RSJ International Conference on Intelligent Robots and Systems (IROS 2003)(Cat. No. 03CH37453)},
  volume={2},
  pages={1044--1049},
  year={2003},
  organization={IEEE}
}

@article{quist2012mechanical,
  title={Mechanical signals at the base of a rat vibrissa: the effect of intrinsic vibrissa curvature and implications for tactile exploration},
  author={Quist, Brian W and Hartmann, Mitra JZ},
  journal={Journal of neurophysiology},
  volume={107},
  number={9},
  pages={2298--2312},
  year={2012},
  publisher={American Physiological Society Bethesda, MD}
}

@article{boubenec2012whisker,
  title={Whisker encoding of mechanical events during active tactile exploration},
  author={Boubenec, Yves and Shulz, Daniel E and Debr{\'e}geas, Georges},
  journal={Frontiers in behavioral neuroscience},
  volume={6},
  pages={74},
  year={2012},
  publisher={Frontiers Media SA}
}

@article{huet2017tactile,
  title={Tactile sensing with whiskers of various shapes: Determining the three-dimensional location of object contact based on mechanical signals at the whisker base},
  author={Huet, Lucie A and Rudnicki, John W and Hartmann, Mitra JZ},
  journal={Soft robotics},
  volume={4},
  number={2},
  pages={88--102},
  year={2017},
  publisher={Mary Ann Liebert, Inc. 140 Huguenot Street, 3rd Floor New Rochelle, NY 10801 USA}
}

@inproceedings{emnett2018novel,
  title={A novel whisker sensor used for 3d contact point determination and contour extraction},
  author={Emnett, Hannah and Graff, Matthew and Hartmann, Mitra},
  booktitle={Robotics Science and Systems},
  volume={14},
  number={June 2018},
  year={2018}
}

@article{sofla2024haptic,
  title={Haptic Localization with a Soft Whisker from Moment Readings at the Base},
  author={Sofla, Mohammad Sheikh and Vayakkattil, Srikishan and Calisti, Marcello},
  journal={Soft Robotics},
  year={2024},
  publisher={Mary Ann Liebert, Inc., publishers 140 Huguenot Street, 3rd Floor New~…}
}

@article{prescott2009whisking,
  title={Whisking with robots},
  author={Prescott, Tony J and Pearson, Martin J and Mitchinson, Ben and Sullivan, J Charles W and Pipe, Anthony G},
  journal={IEEE robotics \& automation magazine},
  volume={16},
  number={3},
  pages={42--50},
  year={2009},
  publisher={IEEE}
}

@article{okamura2009haptic,
  title={Haptic feedback in robot-assisted minimally invasive surgery},
  author={Okamura, Allison M},
  journal={Current opinion in urology},
  volume={19},
  number={1},
  pages={102--107},
  year={2009},
  publisher={LWW}
}

@article{wang2023tactile,
  title={Tactile perception: a biomimetic whisker-based method for clinical gastrointestinal diseases screening},
  author={Wang, Zeyu and Lo, Frank P-W and Huang, Yunran and Chen, Junhong and Calo, James and Chen, Wei and Lo, Benny},
  journal={npj Robotics},
  volume={1},
  number={1},
  pages={3},
  year={2023},
  publisher={Nature Publishing Group UK London}
}

@article{eberhardt2016development,
  title={Development of an artificial sensor for hydrodynamic detection inspired by a seal’s whisker array},
  author={Eberhardt, William C and Wakefield, Brendan F and Murphy, Christin T and Casey, Caroline and Shakhsheer, Yousef and Calhoun, Benton H and Reichmuth, Colleen},
  journal={Bioinspiration \& biomimetics},
  volume={11},
  number={5},
  pages={056011},
  year={2016},
  publisher={IOP Publishing}
}

@article{nguyen2022mechanics,
  title={Mechanics and morphological compensation strategy for trimmed soft whisker sensor},
  author={Nguyen, Nhan Huu and Ho, Van Anh},
  journal={Soft Robotics},
  volume={9},
  number={1},
  pages={135--153},
  year={2022},
  publisher={Mary Ann Liebert, Inc., publishers 140 Huguenot Street, 3rd Floor New~…}
}

@inproceedings{n2010texture,
  title={Texture discrimination with artificial whiskers in the robot-rat Psikharpax},
  author={N’Guyen, Steve and Pirim, Patrick and Meyer, Jean-Arcady},
  booktitle={International Joint Conference on Biomedical Engineering Systems and Technologies},
  pages={252--265},
  year={2010},
  organization={Springer}
}

@article{evans2009spectral,
  title={Spectral template based classification of robotic whisker sensor signals in a floor texture discrimination task},
  author={Evans, Mat and Fox, Charles W and Pearson, Martin J and Prescott, Tony J},
  journal={Proceedings TAROS},
  pages={19--24},
  year={2009}
}

@article{xu2021triboelectric,
  title={A triboelectric-based artificial whisker for reactive obstacle avoidance and local mapping},
  author={Xu, Peng and Wang, Xinyu and Wang, Siyuan and Chen, Tianyu and Liu, Jianhua and Zheng, Jiaxi and Li, Wenxiang and Xu, Minyi and Tao, Jin and Xie, Guangming},
  journal={Research},
  year={2021},
  publisher={AAAS}
}

@incollection{bauchau2009euler,
  title={Euler-Bernoulli beam theory},
  author={Bauchau, Oliver A and Craig, James I},
  booktitle={Structural analysis},
  pages={173--221},
  year={2009},
  publisher={Springer}
}

@inproceedings{stocking2010capacitance,
  title={A capacitance-based whisker-like artificial sensor for fluid motion sensing},
  author={Stocking, JB and Eberhardt, WC and Shakhsheer, YA and Calhoun, BH and Paulus, JR and Appleby, M},
  booktitle={SENSORS, 2010 IEEE},
  pages={2224--2229},
  year={2010},
  organization={IEEE}
}

@article{chen2022comparative,
  title={Comparative study on 3D optical sensors for short range applications},
  author={Chen, Rui and Xu, Jing and Zhang, Song},
  journal={Optics and lasers in engineering},
  volume={149},
  pages={106763},
  year={2022},
  publisher={Elsevier}
}

@inproceedings{lepora2018tacwhiskers,
  title={TacWhiskers: Biomimetic optical tactile whiskered robots},
  author={Lepora, Nathan F and Pearson, Martin and Cramphorn, Luke},
  booktitle={2018 IEEE/RSJ International Conference on Intelligent Robots and Systems (IROS)},
  pages={7628--7634},
  year={2018},
  organization={IEEE}
}

@article{han2016assessing,
  title={Assessing proprioception: a critical review of methods},
  author={Han, Jia and Waddington, Gordon and Adams, Roger and Anson, Judith and Liu, Yu},
  journal={Journal of sport and health science},
  volume={5},
  number={1},
  pages={80--90},
  year={2016},
  publisher={Elsevier}
}

@article{mitchinson2007feedback,
  title={Feedback control in active sensing: rat exploratory whisking is modulated by environmental contact},
  author={Mitchinson, Ben and Martin, Chris J and Grant, Robyn A and Prescott, Tony J},
  journal={Proceedings of the Royal Society B: Biological Sciences},
  volume={274},
  number={1613},
  pages={1035--1041},
  year={2007},
  publisher={The Royal Society London}
}

\end{document}